\documentclass[aps,prd,a4paper,twocolumn,amsmath,showpacs,superscriptaddress,nofootinbib,preprintnumbers]{revtex4-1}

\usepackage{verbatim}
\usepackage[T1]{fontenc}
\usepackage[utf8]{inputenc}
\usepackage[american]{babel}
\usepackage{epsfig}
\usepackage{graphicx}
\usepackage{booktabs}
\usepackage{multirow}
\usepackage{dcolumn}
\usepackage{amsmath}
\usepackage{mathtools}
\usepackage{amsfonts}
\usepackage{amssymb}
\usepackage{epstopdf}
\usepackage{bm}
\usepackage{siunitx}
\usepackage{braket}
\usepackage{enumitem}
\usepackage{soul}
\usepackage[table]{xcolor}
\usepackage{color}
\usepackage{transparent}
\usepackage{comment}
\usepackage{pifont}
\usepackage{soul}

\newcommand{\nuD}{\nu_{\rm dm}}
\newcommand{\nuL}{\nu_\Lambda}
\newcommand{\rhoD}{\rho_{\rm dm}}

\newcommand{\rhoDE}{\rho_{\rm de}}
\newcommand{\w}{w_0}
\newcommand{\wa}{w_a}
\newcommand{\xip}{\xi^{+}}
\newcommand{\xim}{\xi^{-}}
\newcommand{\CC}{\Lambda}
\newcommand{\rv}{\rho_{\rm vac}}
\newcommand{\rvo}{\rho_{\rm vac}^0}

\usepackage{booktabs}
\usepackage{multirow}
\usepackage{dcolumn}
\usepackage{colortbl}

\renewcommand{\arraystretch}{1.5}

\begin{document} 

\title{Dynamical Dark Energy models in light of the latest observations}

\author{Javier de Cruz P\'erez}
\email{jdecruz@ugr.es}
\affiliation{Departamento de Física Te\'orica y del Cosmos, Universidad de Granada, E-18071, Granada, Spain}

\author{Adri\`a G\'omez-Valent}
\email{agomezvalent@icc.ub.edu}
\affiliation{Departament de Física Quàntica i Astrofísica (FQA) and Institut de Ciències del Cosmos (ICCUB), Universitat de Barcelona (UB), c. Martí i Franqués, 1, 08028 Barcelona, Catalonia, Spain}

\author{Joan Sol\`a Peracaula}
\email{sola@fqa.ub.edu}
\affiliation{Departament de Física Quàntica i Astrofísica (FQA) and Institut de Ciències del Cosmos (ICCUB), Universitat de Barcelona (UB), c. Martí i Franqués, 1, 08028 Barcelona, Catalonia, Spain}

\preprint{}
\begin{abstract}
In this paper, we study several models and parameterizations of dynamical dark energy (DE) that have been studied already in the past, in conjunction with the recently proposed model $w$XCDM, the running vacuum model (RVM) with and without a threshold at $z=1$ and two variants of it, the RRVM and the ``flipped RVM'', and compare them all with the concordance $\Lambda$CDM model and the popular $w_0w_a$CDM parameterization.  We use two standard sets of cosmological data, one including distant supernovae from Pantheon$+$ and the other from DES-Y5. The rest of the data (BAO from DESI DR2 and CMB from Planck PR4) are shared by the two sets. They are analyzed using the state-of-the-art techniques. No structure formation data are utilized for this analysis and no use is made of the SH0ES calibration of $H_0$.  Even so, we find that the flipped RVM and to a lesser extent the $w$XCDM and the RVM with threshold, point to significant evidence of dynamical DE, at a level comparable to $w_0w_a$CDM, more conspicuously for the dataset that involves DES-Y5 observations. We also find that while more traditional models studied in the past, in which there is an exchange between vacuum energy and cold dark matter (through e.g. an interactive source  proportional either to the density of dark matter or to that of vacuum) still hint at dynamical DE,  the strength of the statistical signal (which we assess through information criteria and other estimators) is nevertheless less pronounced. Finally, we discuss the ability of the various models to explain the data by performing an analysis of their effective equation-of-state parameters and corresponding evolution of their  dark energy densities.
\end{abstract}

\maketitle


\section{Introduction}\label{sec:intro}
Dark Energy (DE) has been generically invoked as the ultimate cause of the accelerated expansion of the universe. Although we do not yet know what it is,  the first historical proposal for the DE has been the vacuum energy density $\rv=\CC/(8\pi G_N)$ associated with the  cosmological constant, $\Lambda$, in Einstein's equations, where $G_N$ is Newton's constant. Today, $\Lambda$  constitutes a cornerstone of the standard or concordance model of cosmology, aka $\CC$CDM. Despite the latter still being the leading paradigm to account for the cosmological evolution of our universe, in the last few decades it has been pestered with several theoretical problems, which have undermined its consistency and have become a serious handicap to its reputation. These involve not only prime theoretical conundrums, such as the cosmological constant problem (CCP) \cite{Weinberg:1988cp,Sahni:1999gb,Carroll:2000fy,Peebles:2002gy,Padmanabhan:2002ji,Copeland:2006wr,Sola:2013gha,Sola:2015rra,SolaPeracaula:2022hpd}, but also phenomenological pitfalls (which might hide more fundamental problems) still persisting in modern observations, such as the so-called cosmological tensions between the $\CC$CDM predictions and observations. Particularly worrisome is the mismatch between the local measurement of the Hubble-Lema\^itre parameter, $H_0$, and the value determined from CMB observations and the  inverse distance ladder; and also (but to a lesser extent)  the exceeding growth rate of structure formation predicted by the $\CC$CDM at low redshifts; see, e.g. \cite{Perivolaropoulos:2021jda,Abdalla:2022yfr,CosmoVerseNetwork:2025alb} and references therein for a comprehensive review of these tensions and other persistent anomalies. Recently, a new kind of (severe) tension has popped up in the already fairly battered landscape of the $\CC$CDM. It bears relation with the observations of the  James Webb Space Telescope (JWST)\,\cite{Gardner:2006ky,Labbe:2022ahb}. They  have revealed the existence of an unexpectedly large population of supermassive galaxies at large redshifts in the approximate range $z\gtrsim 5-10$, which is completely at odds with the expectations of the $\CC$CDM model. This situation is certainly troubling. All in all, we cannot exclude the need for a new paradigm in the horizon that may  eventually encompass fundamental aspects of the  old one but at the same time be capable of making bold new proposals.

To address these problems efficiently, we need to gather more information on the nature of  dark energy.
Intriguingly enough, recent measurements of baryon acoustic oscillations (BAO) with the Dark Energy Spectroscopic Instrument (DESI) combined with cosmic microwave background (CMB) data and supernovae of Type Ia (SNIa) from various observational teams come to rescue and suggest that the DE could be a dynamical component of the universe\,\cite{DESI:2024mwx,DESI:2025zgx}. These signals are also present when the DESI data are replaced with earlier data from the Sloan Digital Sky Survey (SDSS) \cite{Park:2024vrw, Park:2024pew,Gomez-Valent:2024ejh,Giare:2025pzu,Lu:2025gki}. If confirmed, a rigid $\CC$-term should definitively be ruled out, and this would certainly break with the old paradigm encoded in the $\CC$CDM. In fact, a number of recent analyzes have demonstrated the effectiveness of dynamical DE in improving the description of cosmological observations beyond the standard $\CC$CDM, also in the context of model-agnostic reconstructions of the effective DE properties \cite{DESI:2024aqx,DESI:2025fii,Berti:2025phi,Gonzalez-Fuentes:2025lei}. These studies suggest that the effective DE density underwent a transition from phantom to quintessence around $z\sim 0.5$. The possibility that the DE may be dynamical has a long pedigree. Already about ten years ago, the positive influence of dynamical DE on the quality fit to the overall cosmological data was emphasized by several devoted studies, which analyzed a large set of cosmological observations of various sorts; see, e.g. \cite{Sola:2015wwa,Sola:2016jky,SolaPeracaula:2016qlq,DiValentino:2016hlg,Zhao:2017cud,SolaPeracaula:2017esw,Sola:2017znb}. For older preliminary studies, see  e.g. \cite{Shafieloo:2005nd,Basilakos:2009wi,Sahni:2014ooa,Salvatelli:2014zta,Gomez-Valent:2014rxa,Gomez-Valent:2014fda},  among others. More recent work by different groups has riveted the issue of dynamical DE in the light of the latest observations. An incomplete list of modern references devoted to these studies from different perspectives includes, e.g. \cite{Gomez-Valent:2020mqn,Gomez-Valent:2021cbe,SolaPeracaula:2021gxi,Goh:2022gxo,Maroto:2023toq,Zhai:2023yny,SolaPeracaula:2023swx,Gomez-Valent:2023hov,Akarsu:2023mfb,Jaramillo-Garrido:2023cor,Gomez-Valent:2023uof,Chakraborty:2024xas,Gomez-Valent:2024tdb,Alonso-Lopez:2023hkx, Giare:2024ytc,Giare:2024smz,Park:2024vrw,Wetterich:2024ieb,Toda:2024ncp,Ye:2024ywg,Wolf:2024eph,Jiang:2024xnu,Ye:2024zpk,Gomez-Valent:2024ejh,Tiwari:2024gzo,Odintsov:2024woi,RoyChoudhury:2024wri,Forconi:2025cwp,Khoury:2025txd,Wolf:2025jlc,Sabogal:2025mkp,Giare:2025pzu,Lu:2025gki,Chakraborty:2025syu,Wolf:2025jed,Chen:2025wwn,Giani:2025hhs,Cai:2025mas,Yang:2025mws,Scherer:2025esj,Qin:2025nkk,Asorey:2025hgx,Ozulker:2025ehg,Silva:2025twg,Nojiri:2025low,Braglia:2025gdo,Guedezounme:2025wav,Gohar:2025huk,Rodrigues:2025zvq,Gomez-Valent:2025mfl,Goh:2025upc,Wolf:2025acj,Artola:2025zzb,Li:2025muv,Savaliya:2025cct,Feleppa:2025vop,RoyChoudhury:2025dhe,Carloni:2025dqt,RoyChoudhury:2025iis,Capozziello:2025qmh,Smith:2025uaq,Zhou:2025ugf,Thanankullaphong:2026anl}.

In this paper, we continue our long line of phenomenological investigations on dynamical DE models mentioned above. On the one hand, we revisit interactive DE-DM models of dynamical vacuum energy which have been successfully considered in the past -- with source terms proportional to the dark matter or dark energy densities --, and on the other hand we discuss more recent proposals, such as the $w$XCDM model\,\cite{Gomez-Valent:2024tdb,Gomez-Valent:2024ejh}, which uses the bold notion of ``phantom matter' \cite{Grande:2006nn}  - a concept that must be carefully distinguished from phantom DE -,  together with a new proposal: the ``flipped RVM''. The latter is a variant of the successful running vacuum model \cite{Sola:2015rra,SolaPeracaula:2022hpd}, which we also revisit. Another variant of the RVM (the RRVM\,\cite{SolaPeracaula:2021gxi,SolaPeracaula:2023swx}) is also reexamined. These scenarios provide significant evidence of dynamical DE when confronted with current SNIa, BAO and CMB data. The level of evidence that ensues proves superior to the aforementioned (more traditional) interactive DE-DM models, but  depends on the particular dataset involved. Details are provided in the main body of our study. In addition, all these models are compared with the concordance $\CC$CDM and  with the $w_0w_a$CDM parameterization of the DE, which has been widely used recently by different authors, including the DESI collaboration.

The flow chart of our presentation is the following. In Sec.\,\ref{sec:models} we define the dynamical DE models that are under scrutiny. Except for the most recent ones, which contain special features, we only provide essential background information on the matter and DE densities and refer the reader to the existing literature for more details. 
Sec. \ref{sec:effectiveEoS} describes the construction of an effective DE equation-of-state (EoS) parameter in interacting models.  This kind approach has been used in the past\,\cite{Sola:2005et,Das:2005yj,Sola:2005nh,Basilakos:2013vya} and enables a more direct comparison with previous studies in the literature.  In Sec.\,\ref{sec:perturbations} we discuss our treatment of linear perturbations for the different models, accounting for the dynamical DE effects in the formation of the large-scale structure (LSS) of the universe and including the non-null energy transfer between the components of the dark sector in the interacting models. In Sec.\,\ref{sec:data} we list the cosmological data sources used in our fitting analysis, and in the next section we furnish our numerical results accompanied with the corresponding discussion. Finally, in Sec.\,\ref{sec:conclusions} we deliver our conclusions. 

\section{Dynamical dark energy models}\label{sec:models}

\subsection{RVM and RRVM (Running Vacuum models)}
The RVM has a direct connection with quantum field theory (QFT) in curved spacetime, see e.g. \cite{SolaPeracaula:2022hpd,Moreno-Pulido:2020anb,Moreno-Pulido:2022phq}. \textcolor{black}{In addition, it provides a possible unified theory of dark energy and inflation on fundamental grounds\,\cite{SolaPeracaula:2025yco,SolaPeracaula:2026trz}.} Phenomenologically, it has been analyzed on previous occasions under different conditions and has proved highly competitive with the standard $\CC$CDM. See, among others, the following detailed works\,\cite{Gomez-Valent:2014fda,Sola:2016jky,Gomez-Valent:2014rxa,Sola:2015wwa,Sola:2017znb,SolaPeracaula:2017esw,Sola:2016jky,SolaPeracaula:2016qlq}.  The canonical version of the RVM has a vacuum energy density (VED) evolving with the Hubble rate $H=\dot{a}/a$ as follows, with the dot denoting a derivative with respect to the cosmic time: 
\begin{equation}\label{eq:rvmcanonical}
\rv(H) = \frac{3}{8\pi{G_N}}\left(c_0 + \nu H^2\right)\,,
\end{equation}
in which the dimensionless parameter $\nu$  (the only additional free parameter in the RVM) must be fitted from the cosmological data. For $\nu=0$ we recover the $\CC$CDM, since then $\rv=\CC/(8\pi G_N)$, where $\CC=3 c_0$ is the traditional cosmological constant. For $\nu\neq0$ the vacuum undergoes some mild evolution with the expansion. We expect $|\nu|\ll1$, if the RVM has to remain close to the $\CC$CDM. Theoretically, $\nu$ is connected with the renormalization group running of the VED.  It leads to a quantum correction on top of the classical theory,  so $\nu$ is naturally predicted to be small in absolute value\,\cite{Moreno-Pulido:2020anb,Moreno-Pulido:2022phq}. We note that for  $\nu>0$ (resp.  for $\nu<0$) the vacuum decays into dark matter (DM) particles (resp. DM particles disappear into the vacuum). Interestingly, the ``renormalized $\CC$CDM'' resulting from these quantum corrections in the RVM framework can help to improve the overall fit to the cosmological data as well as to ease the cosmological tensions.

Here, for definiteness, we adopt the setup we used in \cite{SolaPeracaula:2017esw}, in which the vacuum energy only interacts with DM. Baryons and radiation are therefore conserved. We consider pressureless matter and a vacuum EoS parameter $w_{\rm vac}=-1$, so $p_{\rm vac}=-\rho_{\rm vac}$. Solving the model in combination with the Friedmann equations yields the following expressions for the energy densities of dark matter and vacuum as functions of the scale factor:
\begin{equation}\label{eq:rhoDMRRVM}
\rhoD(a)=\rhoD^0\,a^{-3(1-\nu)}+\rho_b^0\left(a^{-3(1-\nu)}-a^{-3}\right)\,,
\end{equation}
and
\begin{equation}\label{eq:rhoLRRVM}
\rv(a)=\rvo+\frac{\nu\rho_m^0}{1-\nu}\left(a^{-3(1-\nu)}-1\right)\,,
\end{equation}
where $\rho_m^0$ and $\rho_b^0$ are today's values of the DM and baryon densities, respectively, and $\rho_m^0=\rho_{\rm dm}^0+\rho_b^0$ is the corresponding total matter energy-density. Since baryons are assumed to be conserved, $\rho_b(a)=\rho_b^0a^{-3}$.
For the sake of simplicity, in the above formulas we have omitted the corrections to these expressions in the radiation-dominated epoch, which have nevertheless been considered in full detail in our numerical analysis and can be found in \cite{SolaPeracaula:2017esw}. Similarly, we also omit in these formulae  the corrections introduced by massive neutrinos, but again have been considered in detail in the numerical analysis, see e.g. \cite{SolaPeracaula:2023swx} for the specific treatment of these contributions.

In the RVM, the ratio $\rho_{\rm vac}(a)/\rho_{\rm dm}(a)\to \nu+\mathcal{O}(\nu^2)$ in the matter-dominated epoch, so there is a tiny but irreducible constant fraction of vacuum energy in that period of the cosmic expansion, which is determined by the small RVM parameter $\nu$. There is therefore, a sort of scaling era, in which the VED evolves in good approximation as the dark matter density. However, this only holds at the level of the energy densities, not the pressures, since the EoS of both species are still those of vacuum and matter, respectively.

The RVM structure can be more general than the canonical one presented before, in that  the VED may evolve not only with $H^2$, but also with $\dot{H}$. When the coefficients of these two homogeneous terms are appropriate, the VED is then proportional to the curvature scalar ${\cal R}=12 H^2+6\dot{H}$. This particular variant of the RVM was called RRVM in \cite{SolaPeracaula:2021gxi,SolaPeracaula:2023swx}. The DE density in this case reads as follows:
\begin{equation}\label{eq:rhov}
\rv(H) = \frac{3}{8\pi{G_N}}\left[c_0 + \nu\left(H^2 + \frac{1}{2}\dot{H}\right)\right].
\end{equation}
Notice that the dynamical term in parenthesis is just ${\cal R}/12$. The advantage of the RRVM behavior over the canonical RVM one is that when we approach and surpass the radiation domain at very high redshifts (beyond the matter-radiation equality point $z_{\rm eq}\simeq 3400$), we have ${\cal R}\propto T^\mu_\mu= \rho-3p=\rho_m\ll \rho_r$ and the VED effects are essentially nullified, thus preserving unscathed the features of the radiation-dominated epoch. The RRVM can also be solved explicitly in terms of the scale factor. The vacuum and DM energy densities take exactly the same form as in Eqs. \eqref{eq:rhoDMRRVM}-\eqref{eq:rhoLRRVM}, but with no corrections in the radiation-dominated epoch for the aforementioned reasons, see \cite{SolaPeracaula:2021gxi,SolaPeracaula:2023swx} for details. We shall also analyze the RRVM here.

\subsection{RVM$_{\text{thr}}$ (RVM with a redshift threshold)}

In this version of the RVM a transition point $z_t$ is introduced in the behavior of the VED, which may be called also a ``redshift threshold''\,\cite{SolaPeracaula:2021gxi,SolaPeracaula:2023swx}. It helps to improve the fit, especially in what concerns the structure formation data (when they are included), since it leads to a suppression of the growth of structure in the low-$z$ universe if $\nu>0$, and it also frees the model from the effects of the VED not only in the radiation-dominated epoch (something that the RRVM variant takes care automatically, see the previous section), but also deep in the matter-dominated era for typical locations of the transition point. The threshold behavior can be expressed in terms of either the scale factor or the redshift. Specifically, if $a>a_t$ (equivalently, $z<z_t$) the energy densities read as in Eqs. \eqref{eq:rhoDMRRVM} and \eqref{eq:rhoLRRVM}. If, instead, $a<a_t$ ($z>z_t$), $\nu=0$ and therefore DM dilutes as in the standard model and  vacuum remains immutable at high redshifts.  This means that  for $a<a_t$  their energy densities read as follows:
\begin{equation}\label{eq:rhoDMRRVM2}
\rhoD(a)=\left[\rhoD^0\,a_t^{3\nu}+\rho_b^0\left(a_t^{3\nu}-1\right)\right]\,a^{-3}\,,
\end{equation}
and 
\begin{equation}\label{eq:rhoLRRVM2}
\rv(a)=\rvo+\frac{\nu\rho_m^0}{1-\nu}\left(a_t^{-3(1-\nu)}-1\right)\,.
\end{equation}
On comparing these expressions with Eqs. \eqref{eq:rhoDMRRVM} and \eqref{eq:rhoLRRVM} we can see that they are continuous at $a=a_t$. Furthermore, the fact that the VED \eqref{eq:rhoLRRVM2} remains frozen at its value at $a=a_t$ for $a<a_t$ is beneficial, as this feature can help improve the fit quality of the RVM with threshold compared to not having it. The reason being that with threshold the VED does not continue to increase (or decrease, depending on the sign of $\nu$) until the last scattering surface. Thus, in the presence of the threshold, $\nu$ can take larger values than in the original RVM or RRVM models without threshold -- and therefore have a stronger impact at low redshifts -- while at the same time avoiding the injection of an exceeding amount of dark energy (vacuum energy, in this case) at the time of decoupling. So, even in the absence of structure formation data, the presence of a threshold can be advantageous. We set it at $a_t=0.5$ ($z_t=1$), as in our previous works \cite{SolaPeracaula:2021gxi,SolaPeracaula:2023swx}. One can check that this is also the typical value obtained when $z_t$ enters the fit.

A redshift threshold can also be introduced for the RRVM variant mentioned in the previous section. We find that the redshift segmentation is completely analogous. The evolution of the cosmological observables of RRVM$_{\rm thr}$ before and after the transition turn out to be very similar to those of RVM$_{\rm thr}$. In addition, we have explicitly verified that the fitting results remain almost identical in both scenarios, as expected in the presence of the threshold. Therefore, we find justified to show here the results obtained for the RVM$_{\rm thr}$ only.

\subsection{$Q_{\rm dm}$ model: interactive source $\propto H\rho_{\rm dm}$}

This model was previously considered in detail in \cite{SolaPeracaula:2017esw,Sola:2017znb} and more recently in \cite{Yang:2025boq,Li:2025muv}. Here the vacuum interacts with a source proportional to the DM density, specifically $Q=3\nu_{\rm dm} H\rho_{\rm dm}$, where $\nu_{\rm dm}$ is a dimensionless parameter to be fitted by comparison with the data. As it may be expected, $\nu_{\rm dm}$ is small ($|\nu_{\rm dm}|\ll1$) and is the analogue of $\nu$ in the RVM, except that model {$Q_{\rm dm}$ is purely phenomenological and in contrast to the RVM it has no known connection with QFT or any other fundamental theory. The densities of DM and vacuum that follow from solving the model read:
\begin{equation}\label{eq:rhodmQdm}
\rho_{\rm dm}(a) = \rhoD^0\,a^{-3(1-\nuD)}\,,
\end{equation}
\begin{equation}\label{eq:rhovacQdm}
\rv(a)=\rvo+\frac{\nuD \rhoD^0}{1-\nuD}\left(a^{-3(1-\nuD)}-1\right)\,.
\end{equation}
Similarly as in the RVM,  for  $\nu_{\rm dm}>0$ (resp.  $\nu_{\rm dm}<0$) the vacuum decays into DM particles (resp. DM particles disappear into the vacuum). Hereafter, for simplicity, the parameter $\nu_{\rm dm}$ will be referred to as $\nu$. We use this simplified notation in the tables.

\subsection{$Q_\Lambda$ model: interactive source $\propto H\rv$}

In this case, the vacuum interaction is with a source which is proportional to the vacuum density itself, namely  $Q=3\nu_{\Lambda} H\rv$, where $\nu_{\Lambda}$ is a free parameter to be fitted by comparison with the data and is numerically small  ($|\nu_\Lambda|\ll1$). This model is also purely phenomenological, with no known connection with a theoretical framework.  This model  was previously analyzed, e.g., in \cite{Sola:2017znb} and more recently in \cite{Zhai:2023yny,Bernui:2023byc,Giare:2024ytc,Giare:2024smz,Brito:2024bhh,Li:2025muv,Yang:2025vnm}. The resulting energy densities are the following:
\begin{equation}
\rhoD(a)=\rhoD^0\,a^{-3}+\frac{\nuL\rvo}{1-\nuL}\left(a^{-3\nuL}-a^{-3}\right)\,,
\end{equation}

\begin{equation}
\rv(a)=\rvo\,a^{-3\nuL}\,.
\end{equation}
Similar considerations follow as before: $\nu_{\Lambda}>0$ or $\nu_\Lambda<0$ imply that the vacuum decays into DM particles or that DM particles disappear into the vacuum, respectively. During the matter-dominated epoch, the VED   evolves essentially in a logarithmic way with the scale factor, that is to say, $\rho_{\rm vac}(a)\simeq 1-3\nu_\Lambda\ln a$, whereas the matter density does not depart significantly from the standard law $\rho_{\rm dm}(a)\sim a^{-3}$. Therefore, in contrast to the situation that we discussed for models with a source  that is proportional to the total matter or DM energy densities (such as the RVM or the $Q_{\rm dm}$), in the $Q_\Lambda$ case the ratio $\rho_{\rm vac}(a)/\rho_m(a)\to 0$ in the matter-dominated epoch}. This is because, by construction, the source itself is very suppressed at high redshifts, where the VED is very small. As a bonus, this model does not inject any significant amount of VED at the decoupling epoch. Similarly as for $\nu_{\rm dm}$, parameter $\nu_{\Lambda}$ will be referred to simply as $\nu$ in the tables.

\subsection{Flipped RVM}\label{subsec:flipped}

This scenario is a variant of the original RVM that we propose here for the first time. As we shall see, it renders a significant phenomenological output. The model has  a double threshold at redshifts $z_1$ and $z_2$ ($>z_1$), \textcolor{black}{with the important difference that the $\nu$ parameter of the RVM  performs a sign flip at $z_1$ and the vacuum dynamics is switched off for $z>z_2$. Despite the fact that the motivation for considering a double threshold is essentially phenomenological in the present analysis, it proves quite effective and produces a competitive fit. In addition, since in this model the parameter $\nu$ of the RVM performs a sign flip (therein its name), it could implement in an effective way the possibility of having DE with negative energy density and positive pressure in some redshift interval (an option which is called `phantom matter' \cite{Grande:2006nn}). This option was demonstrated to be phenomenologically fruitful in recent works by two of us\,\cite{Gomez-Valent:2024ejh,Gomez-Valent:2024tdb} and we revisit it in Sec. \ref{sec:PM}. Here we only mention it in that the flipped RVM model under appropriate data could mimick phantom matter behavior too.  The possibility of having `bubbles of phantom matter' in the universe has recently been addressed in  theoretical works within the context of stringy versions of the RVM, see e.g. \cite{Mavromatos:2021urx,Mavromatos:2020kzj}.} 

In this flipped scenario, the VED can experience a transition from a growing to a decreasing evolution with the expansion (if $\nu>0$) or the other way around (if $\nu<0$), hopefully mimicking the behavior that is preferred by current cosmological observations in the context of self-conserved DE models, with a crossing of the phantom divide from phantom to quintessence at $z\sim 0.5$, as explained in the Introduction. One cannot exclude a priori that the flipping mechanism may have a preference for choosing negative VED values in some redshift range, depending on whether this feature  leads to a better overall fit to the data being used here. If the flipped RVM is able to reproduce that behavior or not is something that is not trivial to say at this point and can only be elucidated after we perform the numerical analysis. Furthermore, since the flipped RVM is an interacting model, it will be convenient to study its effective equation of state, using the tools that will be described in Sec. \ref{sec:effectiveEoS}.

For convenience, let us define 
 $\xi^{\pm}=1\pm \nu$, where $\nu$ is the single parameter of the RVM. The structure of the flipped RVM is inherited from that of the RVM, but allowing for a sign flip in the value of the $\nu$ parameter in an intermediate redshift range. This generates a counter effect in the evolution of the DE, which transits from growing ($\nu<0$)  to decreasing ($\nu>0$) behavior with the expansion.  Specifically, the segmentation of the DE (and corresponding matter density) throughout cosmic history is the following. First, in the lowest redshift range, near our time, $0<z<z_1$ ($1>a>a_1$),  we have
\begin{equation}
\rho_m(a) = \rho_m^0\,a^{-3\xim}
\end{equation}
and
\begin{equation}
\rv(a) = \rvo+\rho_m^0\left(\frac{1}{\xim}-1\right)\left(a^{-3\xim}-1\right)\,.
\end{equation}
Then, in the intermediate redshift range $z_1<z<z_2$ ($a_1>a>a_2$), the $\nu$ parameter flips sign and the energy densities read as follows:
\begin{equation}
 \rho_m(a) = \rho_m^0\,a_1^{3(\xip-\xim)}a^{-3\xip}   
\end{equation}
and
\begin{eqnarray}
\rv(a)&=&\rvo+\rho_m^0\left(\frac{1}{\xim}-\frac{1}{\xip}\right)a_1^{-3\xim}+\\
&&\rho_m^0\left(1-\frac{1}{\xim}\right)+\rho_m^0\left(\frac{1}{\xip}-1\right)a_1^{3(\xip-\xim)}a^{-3\xip}\,.\nonumber
\end{eqnarray}
Finally, in the high redshift range $z>z_2$ ($a<a_2$) the DM density evolves as $a^{-3}$ and the VED remains constant:

\begin{equation}
\rho_m(a)=\rho_m^0\,a_1^{3(\xip-\xim)}a_2^{3(1-\xip)}a^{-3}
\end{equation}
and
\begin{eqnarray}
    \rv(a)&=&\rvo+\rho_m^0\left(\frac{1}{\xim}-\frac{1}{\xip}\right)a_1^{-3\xim}\\
    &&+\rho_m^0\left(1-\frac{1}{\xim}\right)+\rho_m^0\left(\frac{1}{\xip}-1\right)a_1^{3(\xip-\xim)}a_2^{-3\xip}\,.\nonumber
\end{eqnarray}
Notice that these densities are continuous functions in the entire redshift range. As can be seen, the flippped RVM has three additional independent free parameters compared to the $\CC$CDM, namely $(\nu, z_1, z_2)$.  For $\xi^+=\xi^-=1$ 
(i.e. for $\nu=0$) we recover the $\CC$CDM.

\subsection{$w$XCDM (phantom matter proposal)}\label{sec:PM}

This model was originally put forward in a recent work \cite{Gomez-Valent:2024tdb}. It is based on the possible existence of a peculiar form of DE  which carries a negative energy density but a positive pressure. It satisfies the strong energy condition, as usual matter, and for this reason it was called ``phantom matter'' in \cite{Grande:2006nn}. Such a form of DE, which is in the antipodes of the usual phantom DE region in the EoS diagram of the cosmological fluids (see Fig. 1 of \cite{Gomez-Valent:2024tdb}),  could help explaining the unexpected large population of supermassive galaxies at large redshifts, as observed by the JWST (see the Introduction). In addition, the model has proven to render an outstanding fit to the cosmological data, together with a very positive impact on the cosmological tensions when transversal (i.e. 2D) BAO is used \cite{Gomez-Valent:2024tdb,Gomez-Valent:2024ejh}. With BAO 3D, however, the impact on the tensions is significantly lessened. Theoretical models justifying the possible existence of phantom matter can be found in the literature and are based once more on stringy formulations of the RVM\,\cite{Mavromatos:2021urx,Mavromatos:2020kzj}.

For the $w$XCDM, the segmentation of DE reads as follows. For $a>a_t$ ($z<z_t$):
\begin{equation}
\rhoDE(a)=\rhoDE^0\,a^{-3(1+w_Y)}\,.
\end{equation}
In this low-$z$ segment the component $Y$ behaves as quintessence (as found by direct fit to the data).
In contrast, for $a<a_t$ ($z>z_t$):
\begin{equation}
\rhoDE(a)=-\rhoDE^0 a_t^{3(w_X-w_Y)}a^{-3(1+w_X)} 
\end{equation}
and in this higher $z$ domain the DE can behave as phantom matter. As in \cite{Gomez-Valent:2024ejh}, we use the prior $w_X\in [-2,-0.5]$.  The additional free parameters of the $w$XCDM compared to the $\CC$CDM are $(z_t, w_X,w_Y)$. This model has the same number of parameters as the flipped RVM. Note that the particular case $w_X=w_Y=-1$ renders the so-called $\CC_s$CDM model \cite{Akarsu:2023mfb}. This model was thoroughly analyzed in conjunction with the $w$XCDM in \cite{Gomez-Valent:2024tdb,Gomez-Valent:2024ejh}. In the current reanalysis we shall address the $w$XCDM only since it encompasses the $\CC_s$CDM as a particular case and hence it generally provides a better fit.

\subsection{$w_0w_a$CDM (also denoted CPL) parameterization}

This parameterization of the DE  was originally introduced in \cite{Chevallier:2000qy,Linder:2002et}. It is based on the following dynamical form of the EoS of the DE:
\begin{equation}\label{eq:CPL}
 w(z)=w_0+w_a(1-a)=w_0+\frac{w_a z}{1+z}\,,
\end{equation}
which we have expressed both in terms of the scale factor and the redshift ($1+z=1/a$).
A particular case is the $w$CDM parameterization\,\cite{Turner:1997npq}, in which $w_0$ is simply called $w$ and  $w_a=0$. The EoS in this simpler case is non-dynamical (although the DE is still evolving with the expansion).  We shall not consider it, since it fails in describing the cosmological data as efficiently as the CPL, as shown in \cite{Gomez-Valent:2024ejh,DESI:2025zgx,Gonzalez-Fuentes:2025lei}.
The DE density associated with the CPL parameterization can be computed easily:
\begin{equation}
\rhoDE(a)=\rhoDE^0\,a^{-3(1+\w+\wa)}e^{-3\wa(1-a)}\,.
\end{equation}
The CPL parametrization is quite flexible. Depending on the values of $w_0$ and $w_a$, it can mimic a large family of freezing and thawing quintessence models \cite{dePutter:2008wt} (if $w_0>-1$ and $w_a>-1-w_0$), phantom DE models (if $w_0<-1$ and $w_a<-1-w_0$), as well as quintom scenarios \cite{Cai:2009zp} and, in particular, quintom-B scenarios with a transition from phantom DE at large redshift to quintessence at small redshift (if $w_0>-1$ and $w_a<-1-w_0$) as those favored by current observations \cite{DESI:2025zgx,DESI:2025fii,Gonzalez-Fuentes:2025lei,Keeley:2025rlg,Ozulker:2025ehg}. The opposite behavior, with a transition from quintessence to phantom DE exhibited in quintom-A models, is realized if $w_0<-1$ and $w_a>-1-w_0$. It is therefore useful to take the CPL parameterization as a benchmark to analyze the relative performance of the various dynamical dark energy models under study.


\section{Effective equation of state of dark energy}\label{sec:effectiveEoS}

In all of the models described in the previous section that include an interaction between vacuum and dark matter, the vacuum has an EoS parameter $w_{\rm vac}=-1$. However, standard model-agnostic reconstructions of the background DE properties are typically performed under the assumption of a composite system consisting of self-conserved pressureless matter and dark energy, with no direct cross-talk between each other, see, e.g., \cite{DESI:2024aqx,DESI:2025fii,Berti:2025phi,Gonzalez-Fuentes:2025lei}. In these reconstructions, the DE should be regarded as an effective self-conserved component, despite that the original DE may not be conserved (e.g. it may interact with dark matter).
Nevertheless, the effective description provides a convenient parametrization of various possible sources of new physics beyond $\Lambda$CDM -- arising, for example, from modified gravity or non-standard interactions within the dark sector. Such new physics are encapsulated at the background level through the effective dark energy EoS parameter, $w_{\rm eff,de}(z)$, which in general can  be different from $-1$, and in fact take a non-trivial dependence on the redshift $z$.

It would be very useful to compare the effective dark energy EoS parameter associated to the various interacting models presented in Sec. \ref{sec:models} with the one extracted from the analysis of current cosmological data applying model-independent methods, see, e.g. \cite{DESI:2025fii,Berti:2025phi,Gonzalez-Fuentes:2025lei}. This can be done quite  straightforwardly following the approach of the well-known old references \cite{Sola:2005et,Das:2005yj,Sola:2005nh,Basilakos:2013vya}. With the aid of an auxiliary self-conserved pressureless matter component with energy density $\tilde{\rho}_m(a)=\tilde{\rho}_m^0a^{-3}$ it is possible to express the total energy density and pressure of the interacting models as follows: 

\begin{equation}
 \rho_{\rm tot}(a)=\rho_{\rm vac}(a)+\rho_m(a)\equiv\tilde{\rho}_{\rm de}(a)+\tilde{\rho}_m(a)\,,   
\end{equation}

\begin{equation}
p_{\rm tot}(a)=p_{\rm vac}(a)\equiv\tilde{p}_{\rm de}(a)\,,    
\end{equation}
where $\tilde{\rho}_{\rm de}$ and $\tilde{p}_{\rm de}$ are the effective (self-conserved) DE density and pressure, respectively, and, for simplicity, here we neglect the contribution of radiation, since we are only interested in the shape of $w_{\rm eff,de}(a)$ in           the late universe, at $z\lesssim 2.5$. Hence, we have,

\begin{equation}\label{eq:effectiveDens}
\tilde{\rho}_{\rm de}(a)=\rho_{\rm vac}(a)+\rho_m(a)-\tilde{\rho}_m(a)\,,
\end{equation}

\begin{equation}
\tilde{p}_{\rm de}(a) = -\rho_{\rm vac}(a)\,,
\end{equation}
and the effective EoS parameter reads, 

\begin{eqnarray}\label{eq:effectiveEoS}
w_{\rm eff,de}(a)&=&\frac{\tilde{p}_{\rm de}(a)}{\tilde{\rho}_{\rm de}(a)}\\&=&-1+\frac{\rho_m(a)-\tilde{\rho}_m(a)}{\rho_m(a)+\rho_{\rm vac}(a)-\tilde{\rho}_m(a)}\,.\nonumber    
\end{eqnarray}
The explicit formula of $w_{\rm eff,de}$ in terms of the scale factor (or the redshift) can be easily obtained making use of the expressions of the energy densities displayed in Sec. \ref{sec:models}. It is important to note that the shape of $w_{\rm eff,de}(a)$ depends on the value of the constant $\tilde{\rho}_m^0$, which can, in principle, be arbitrary, since by construction any choice of $\tilde{\rho}_m^0$ yields exactly the same background cosmology as that of the original interacting model. In this paper, however, we adopt a natural and physically motivated choice: the value of $\tilde{\rho}_m^0$ inferred from the model-agnostic reconstructions of Ref.~\cite{Gonzalez-Fuentes:2025lei} (see Table~6 therein). This value is, in fact, very close to that obtained using the CPL parametrization, which corresponds to the model yielding the lowest $\chi^2_{\rm min}$ among those analyzed in this work, cf. Sec. \ref{sec:perturbations}.

Another aspect worth discussing is that there are two factors that can cause $w_{\rm eff,de}(a)$ to depart from $-1$: the presence of DE dynamics, and a difference between $\rho_m^0$ and $\tilde{\rho}_m^0$. These two conditions are distinct, and both contribute to shaping the form of $w_{\rm eff,de}(a)$.
 
The analysis of $w_{\rm eff,de}(a)$ will allow us to examine, among other aspects, whether the data favor an effective crossing of the phantom divide in the interacting models under study -- an important feature of dynamical dark energy scenarios consistent with current CMB+BAO+SNIa observations \cite{DESI:2025zgx,DESI:2025fii,Gonzalez-Fuentes:2025lei,Keeley:2025rlg,Ozulker:2025ehg}.


\begin{table*}[t!]
\centering
\begin{tabular}{|c ||c |c | c | c| c| }
 \multicolumn{1}{c}{} & \multicolumn{1}{c}{} & \multicolumn{1}{c}{} & \multicolumn{1}{c}{} & \multicolumn{1}{c}{} & \multicolumn{1}{c}{}
\\\hline
{\small Parameter} & {\small $\Lambda$CDM} & {\small CPL} & {\small $Q_{\rm dm}$}& {\small $Q_{\Lambda}$} & {\small $w$XCDM}
\\\hline
$10^2\omega_b$ &  $2.231^{+0.010}_{-0.009}$ & $2.233^{+0.013}_{-0.014}$ & $2.214\pm 0.015$ & $2.232\pm 0.013$ & $2.225\pm 0.014$ \\\hline
$10\,\omega_{\rm dm}$ & $1.178\pm 0.006$ & $1.189\pm 0.008$ & $1.176\pm 0.006$ & $1.212\pm 0.066$ & $1.187\pm 0.010$ \\\hline
$\ln(10^{10}A_s)$ & $3.047^{+0.013}_{-0.014}$ & $3.038^{+0.013}_{-0.014}$ & $3.040\pm 0.014$ & $3.048\pm 0.014$ & $3.041\pm 0.014$ \\\hline
$n_s$ & $0.968^{+0.003}_{-0.004}$ & $0.965\pm 0.004$ & $0.965\pm 0.004$ & $0.968\pm 0.004$  & $0.966\pm 0.004$\\\hline
$\tau$ & $0.059^{+0.009}_{-0.008}$ & $0.053^{+0.007}_{-0.006}$ & $0.054\pm 0.007$ & $0.060^{+0.007}_{-0.008}$ & $0.055\pm 0.007$ \\\hline
$H_0$ [km/s/Mpc] & $68.12\pm 0.27$ & $67.52^{+0.59}_{-0.60}$ & $68.72\pm 0.40$ & $67.86\pm 0.55$  & $67.83\pm 0.56$ \\\hline
$\nu$ & $-$ & $-$ & $-0.00073\pm 0.00034$ & $0.012\pm 0.022$ & $-$\\\hline

$w_0$ & $-$ & $-0.842^{+0.055}_{-0.056}$ & $-$ & $-$ & $-$\\\hline
$w_a$ & $-$ & $-0.59^{+0.22}_{-0.19}$ & $-$ & $-$  & $-$ \\\hline
$w_X$ & $-$ & $-$ & $-$ & $-$  & uncons. \\\hline
$w_Y$ & $-$ & $-$ & $-$ & $-$  & $-0.970^{+0.024}_{-0.025}$ \\\hline
$z_t$ & $-$ & $-$ & $-$ & $-$ & $3.55^{+0.04}_{-1.24}$ \\\hline\hline
$M$ [mag] & $-19.420^{+0.009}_{-0.008}$ & $-19.419^{+0.014}_{-0.015}$ & $-19.403\pm 0.012$ & $-19.426\pm 0.014$ & $-19.423\pm 0.014$\\\hline
$\Omega_m$ & $0.303^{+0.004}_{-0.003}$ & $0.311\pm 0.006$ & $0.297\pm 0.005$ & $0.313\pm 0.019$ & $0.308^{+0.005}_{-0.006}$ \\\hline
$\sigma_{12}$ & $0.794^{+0.005}_{-0.006}$ & $0.801^{+0.006}_{-0.007}$ & $0.805\pm 0.008$ & $0.778^{+0.029}_{-0.037}$ & $0.794^{+0.006}_{-0.007}$ \\\hline
$S_{8}$ & $0.810\pm 0.008$ & $0.822\pm 0.009$ & $0.819\pm 0.009$ & $0.803^{+0.015}_{-0.017}$ & $0.813\pm 0.008$ \\\hline\hline
$\chi^2_{\rm min}$ & $12391.66$ & $12384.08$ & $12388.11$ & $12391.66$ & $12387.29$ \\\hline 
$\Delta \text{AIC}$ & $-$ & $+3.58 $& $+1.55$ & $-2.00$ & $-1.63$ \\\hline
$\Delta \text{DIC}$ & $-$ & $+5.72 $& $+2.72$ & $-0.72$ & $-$ \\\hline
$E_{\Lambda {\rm CDM}}$ &  $-$ & $2.28\sigma$ &  $1.88\sigma$ &  $0.00\sigma$ & $-$ 
\\\hline
\end{tabular}
\caption{Mean values and uncertainties at $68\%$ CL for some of the models under study, obtained using CMB+PantheonPlus+DESI. The value of $\nu$ indicated in the table corresponds to $\nu_{\rm dm}$ or $\nu_{\Lambda}$ according to the model considered. The results of the running vacuum models obtained with the same dataset are displayed in Table \ref{tab:table2}. }\label{tab:table1} 
\end{table*}

\section{Linear Perturbations}\label{sec:perturbations}

We implement the linear perturbation equations in the Einstein-Boltzmann code \texttt{CLASS} \cite{Lesgourgues:2011re,Blas:2011rf} using the synchronous gauge. The perturbed (flat three-dimensional) FLRW metric in the conformal frame reads \cite{Ma:1995ey},
\begin{equation}\label{eq:LineElementSyn}
ds^2=a^2(\tau)[-d\tau^2+(\delta_{ij}+h_{ij})dx^idx^j]\,,
\end{equation}
with $\tau$ the conformal time and
\begin{eqnarray}\label{eq:hFourier}
h_{ij}(\tau,\vec{x})&=&\int d^3k\, e^{-i\vec{k}\cdot\vec{x}}\left[\hat{k}_i\hat{k}_j h(\tau,\vec{k})\right.\nonumber\\
&+&\left.\left(\hat{k}_i\hat{k}_j-\frac{\delta_{ij}}{3} \right)6\eta(\tau,\vec{k})\right]\,,    
\end{eqnarray}
with $\hat{k}_i=k_i/k$. The two fields $h(\tau,\vec{k})$ and $\eta(\tau,\vec{k})$ parameterize the trace and traceless parts of the perturbed metric, respectively.
The perturbed Einstein equations in Fourier space adopt exactly the same form as in the $\Lambda$CDM:

\begin{equation}\label{eq:PerturbTypeIEq1}
\mathcal{H}h^\prime-2\eta k^2=8\pi {G_N}a^2\sum_l\delta\rho_l\,,
\end{equation}
\begin{equation}\label{eq:PerturbTypeIEq2}
\eta^\prime k^2=4\pi {G_N} a^2\sum_l (\bar{\rho}_l+\bar{p}_l)\theta_l\,,
\end{equation}
\begin{equation}\label{eq:PerturbTypeIEq3}
h^{\prime\prime}+2\mathcal{H}h^\prime-2\eta k^2=-24\pi {G_N}a^2\sum_l\delta p_l\,.
\end{equation}
\begin{equation}\label{eq:PerturbTypeIEq4}
h^{\prime\prime}+6\eta^{\prime\prime}+2\mathcal{H}(h^\prime+6\eta^\prime)-2k^2\eta=-24\pi G_N a^2(\bar{\rho}+\bar{p})\sigma\,,
\end{equation}
where $\mathcal{H}\equiv aH$ is the Hubble rate in conformal time, and the sums run over the different matter components. The primes denote derivatives with respect to the conformal time, and $(\bar{\rho}+\bar{p})\sigma$ encodes the information of the anisotropic stress. The bars in these equations indicate background quantities and $\theta_l$ is the divergence of the perturbed 3-velocity of the fluid $l$. 

For the CPL and $w$XCDM models, we assume that whatever constitutes the DE sector does not cluster efficiently, and we simply set the DE sound speed $c_s^2=1$\footnote{\textcolor{black}{The dark energy sound speed in late-time dynamical DE models can only have an impact on the low CMB multipoles, since in this paper we do not use LSS data. The changes induced in these small multipoles remain in any case much below the cosmic variance threshold, even if we take the lowest possible sound speed, $c_s^2=0$ \cite{dePutter:2010vy}. Therefore, with our dataset we cannot distinguish efficiently between late-time DE models with different values of $c_s^2$.}}. In these models, DE influence the large-scale structure history only through its impact on the background functions. In the CPL, we use the parametrized post-Friedmann approach to avoid the problematic singularities at the crossing of the phantom divide \cite{Fang:2008sn}. In the $w$XCDM, while the $Y$ component behaves as an ordinary DE fluid -- which can be quintessence-like or phantom-like depending on the sign of $1+w_Y$ --, $X$ mimics phantom matter and, therefore, the latter can enhance the clustering before the transition, at $z>z_t$, when $X$ starts to compete with pressureless matter. For a more detailed explanation of the effect of the components $Y$ and $X$ on the LSS formation history in the $w$XCDM model, we refer the reader to Refs. \cite{Gomez-Valent:2024tdb,Gomez-Valent:2024ejh}.

The perturbed conservation equations for the other cosmic species are the same as in the standard model. However, in those models with an interaction between dark matter and the vacuum, the conservation equations for these two components change. We consider a vacuum–geodesic CDM interaction such that there is no net momentum transfer between the vacuum and cold dark matter \cite{Wang:2013qy,Wang:2014xca,Gomez-Valent:2018nib,SolaPeracaula:2023swx}, which allows us to fix the gauge by setting $\theta_{\textrm{dm}}=0$ -- as it is done in the $\Lambda$CDM, CPL and $w$XCDM -- and also to set $\delta\rho_{\rm vac}=0$, yielding the following conservation equation,

\begin{equation}\label{eq:pert}
\delta_{\textrm{dm}}^\prime+\frac{h^\prime}{2}-\frac{\bar{\rho}_{\rm vac}^\prime}{\bar{\rho}_{\textrm{dm}}}\delta_{\textrm{dm}}=0\,,
\end{equation}
with $\delta_{\textrm{dm}}=\delta\rho_{\textrm{dm}}/\bar{\rho}_{\textrm{dm}}$ the CDM density contrast. This is actually the only perturbation equation that must be modified in \texttt{CLASS} in order to accommodate the dynamical character of the VED in our framework. The formula of the ratio $\bar{\rho}_{\rm vac}^\prime/\rho_{\rm dm}$ is easily computable in each model as a function of the scale factor using the analytical expressions shown above. If $\bar{\rho}_{\rm vac}^\prime= 0$, i.e., if there is no exchange of energy in the dark sector, we retrieve the CDM equation that is found in the $\Lambda$CDM, as expected.

In this work, we consider adiabatic perturbations for the various matter and radiation species.


\begin{table*}[t!]
\centering
\begin{tabular}{|c ||c |c | c | c|   }
 \multicolumn{1}{c}{} & \multicolumn{1}{c}{} & \multicolumn{1}{c}{} & \multicolumn{1}{c}{} & \multicolumn{1}{c}{}
\\\hline
{\small Parameter} & {\small $\Lambda$CDM}  & {\small Flipped RVM} & {\small RRVM} & {\small RVM$_{\text{thr}}$}
\\\hline
$10^2\omega_b$ &  $2.231^{+0.010}_{-0.009}$  & $2.218\pm 0.014$ & $2.214\pm 0.015$ &  $2.234\pm 0.013$ \\\hline
$10\,\omega_{\rm dm}$ & $1.178\pm 0.006$ & $1.274^{+0.046}_{-0.063}$ & $1.176\pm 0.007$  & $1.263\pm 0.065$\\\hline
$\ln(10^{10}A_s)$ & $3.047^{+0.013}_{-0.014}$  & $3.041\pm 0.014$ & $3.041\pm 0.014$ & $3.050\pm 0.014$\\\hline
$n_s$ & $0.968^{+0.003}_{-0.004}$  & $0.964\pm 0.004$ & $0.964\pm 0.004$  & $0.9687\pm 0.0034$ \\\hline
$\tau$ & $0.059^{+0.009}_{-0.008}$  & $0.054\pm 0.007$ & $0.054\pm 0.007$ & $0.061^{+0.007}_{-0.008}$  \\\hline
$H_0$ [km/s/Mpc] & $68.12\pm 0.27$   & $68.00\pm 0.51$ & $68.72\pm 0.41$ & $67.45\pm 0.58$ \\\hline
$\nu$ & $-$ & $0.029^{+0.010}_{-0.019}$ & $-0.00062\pm 0.00029$ &  $0.029\pm 0.022$ \\\hline
$z_1$ &  $-$  & $2.58\pm 0.84$ & $-$ &  $-$ \\\hline
$z_2$ & $-$ & $6.1^{+2.9}_{-1.8}$ & $-$ &  $-$ \\\hline\hline
$M$ [mag] & $-19.420^{+0.009}_{-0.008}$ & $-19.418\pm 0.013$ & $-19.402\pm 0.012$ & $-19.434\pm 0.014$ \\\hline
$\Omega_m$ & $0.303^{+0.004}_{-0.003}$  & $0.325^{+0.012}_{-0.018}$ & $0.297 \pm 0.005$ & $0.328^{+0.018}_{-0.021}$ \\\hline
$\sigma_{12}$ & $0.794^{+0.005}_{-0.006}$ & $0.759^{+0.026}_{-0.019}$ & $0.806\pm 0.008$ &  $0.755^{+0.028}_{-0.031}$ \\\hline
$S_{8}$ & $0.810\pm 0.008$  & $0.800^{+0.012}_{-0.010}$ & $0.819\pm 0.009$ & $0.795\pm 0.014$ \\\hline\hline
$\chi^2_{\rm min}$ & $12391.66$ & $12383.98$ & $12388.45$ &  $12390.09$ \\\hline
$\Delta\text{AIC}$ & $-$ &$+1.68$ & $+1.21$ &  $-0.43$ \\\hline
$\Delta\text{DIC}$ & $-$ &$-$ & $+2.54$ &  $+0.20$ \\\hline
$E_{\Lambda {\rm CDM}}$ &  - & $1.93\sigma$  & $1.79\sigma$ &  $1.25\sigma$
\\\hline
\end{tabular}                        
\caption{Mean values and uncertainties at $68\%$ CL for the various running vacuum  models under study, obtained using CMB+PantheonPlus+DESI. Those for $\Lambda$CDM are also shown for comparison. }\label{tab:table2} 
\end{table*}

\section{Data and methodology}\label{sec:data}

As explained in the Introduction, the main goal of this work is to assess the evidence for dynamical dark energy from the perspective of the cosmological models described in Sec. \ref{sec:models} and in the light of the current data on CMB, BAO and SNIa. Hence, we use the following individual datasets:

\begin{itemize}
\item \textbf{CMB:} We make use of the \textit{Planck} measurements of the CMB temperature (TT), polarization (EE), and cross-correlation (TE) power spectra. Specifically, we adopt the \texttt{simall}, \texttt{Commander}, and \texttt{NPIPE PR4} \texttt{CamSpec} likelihoods \cite{Efstathiou:2019mdh,Rosenberg:2022sdy} for multipoles $\ell < 30$ and $\ell \geq 30$, respectively, in combination with the \texttt{NPIPE PR4} CMB lensing likelihood \cite{Carron:2022eyg}.

\item \textbf{BAO:} We employ the baryon acoustic oscillation measurements from the DESI Data Release 2 (DR2), taking into account the reported correlations as detailed in Table~IV of Ref.~\cite{DESI:2025zgx}.

\item \textbf{SNIa:} We present results obtained using Pantheon$+$ \cite{Scolnic:2021amr} and DES-Y5 \cite{DES:2024hip,DES:2024jxu} separately, without using the SNIa absolute magnitude calibration from SH0ES. Both in the context of the CPL parametrization and in model-independent dark energy reconstructions, the DES-Y5 dataset yields stronger evidence for dynamical dark energy than Pantheon$+$ \cite{DESI:2024mwx,DESI:2025zgx,Gonzalez-Fuentes:2025lei}. Several studies have pointed to a possible mismatch between the photometry of nearby ($z \lesssim 0.1$) and higher-redshift SNIa in DES-Y5 \cite{Gialamas:2024lyw,Efstathiou:2024xcq,Huang:2025som}, which could partly explain the observed discrepancies. The issue, however, remains under debate \cite{DES:2025tir}. For this reason, we regard it as more reliable to analyze and discuss results obtained with both the Pantheon$+$ and DES-Y5 samples\footnote{Our results were generated before the DES re-analysis \cite{DES:2025sig}, which applies an updated SNIa calibration and obtains evidence for evolving DE with a statistical significance lying in between that found using Pantheon+ and that of the original DES-Y5 sample.}.

\end{itemize}

Therefore, we perform fitting analyses of all the models listed in Sec. \ref{sec:models} using these two cosmological datasets: CMB+PantheonPlus+DESI and CMB+DES-Y5+DESI. These datasets have become kind of standard and will ease a fair comparison of our results with others available in the literature. The inclusion of other datasets, as cosmic chronometers or redshift-space distortions, is avoided in this work precisely for this reason. However,  as shown in previous analyses, they are important and certainly deserve further attention in the future.

As already mentioned in the previous section, we implement all our models in \texttt{CLASS} \cite{Lesgourgues:2011re,Blas:2011rf} and perform the Monte Carlo runs  with the aid of \texttt{Cobaya} \cite{Torrado:2020dgo}. We stop the runs when the Gelman-Rubin criterion \cite{GelmanRubin} $R-1\lesssim 0.02$, and process the resulting chains with the \texttt{Python} package \texttt{GetDist} \cite{Lewis:2019xzd}. 

We compare the fitting performance of the various models using two alternative information criteria -- the Akaike (AIC) \cite{Akaike} and deviance (DIC) \cite{DIC} information criteria -- as well as the likelihood-ratio test \cite{LRtest}. The AIC reads,

\begin{equation}\label{eq:AIC}
{\rm AIC} = \chi^2_{\rm min} + 2n\,,
\end{equation}
where $\chi^2_\mathrm{min}$ is the minimum value of $\chi^2$ and $n$ the total number of parameters of the model. \textcolor{black}{We make use of two minimization algorithms already implemented in the \texttt{Cobaya} framework: \texttt{bobyqa} and the minimizer provided by the \texttt{scipy} package. We carefully tune the minimizer settings in order to robustly reach the best-fit points for each dataset and model considered.}

Large values of AIC -- induced by large values of $\chi^2_\mathrm{min}$ signaling a poor fit to the data and/or the presence of many free parameters indicating a greater model complexity -- translate into a less favored model. The DIC, instead, is defined as

\begin{figure*}
    \centering
    \includegraphics[scale=0.55]{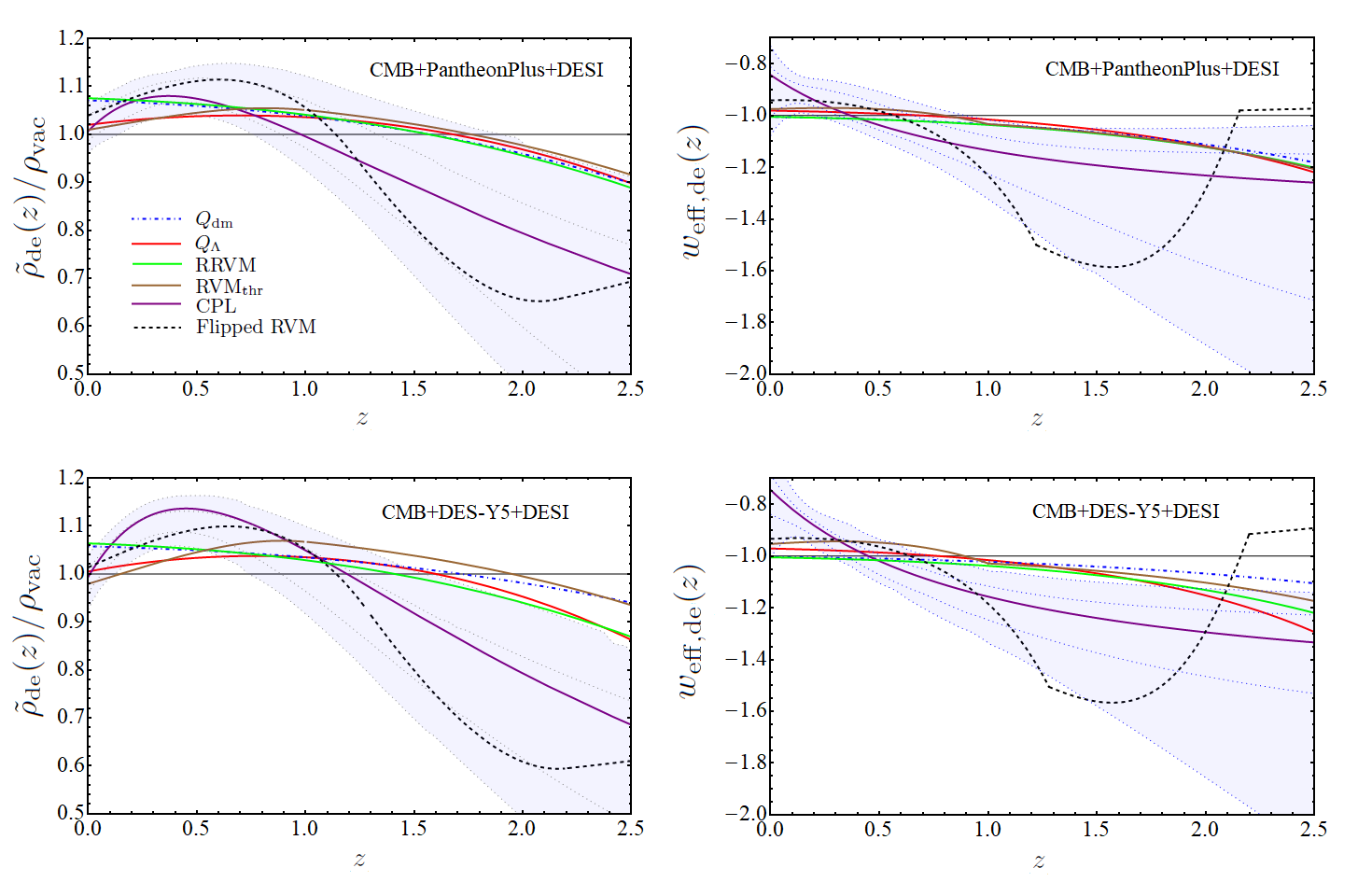}
    \caption{The effective DE density, $\tilde{\rho}_{\rm de}(z)$, and EoS parameter, $w_{\rm eff,de}(z)$, are shown as defined in Eqs. \eqref{eq:effectiveDens} and \eqref{eq:effectiveEoS}, respectively, using the best-fit values of the various models under study obtained in the analyses with CMB+PantheonPlus+DESI (upper row) and CMB+DES-Y5+DESI (lower row). We normalize the effective DE density to the energy density associated with the cosmological constant in the $\Lambda$CDM model, $\rv=\CC/(8\pi G_N)$, as inferred from Planck PR4 data \cite{Rosenberg:2022sdy}. We also display, in light blue, the $68\%$ and $95\%$ CL bands obtained in \cite{Gonzalez-Fuentes:2025lei} using a model-independent approach, the so-called Weighted Function Regression method \cite{Gomez-Valent:2018hwc,Gomez-Valent:2018gvm}. Note that most of the considered DE models, except $Q_{\rm dm}$ and RRVM (without threshold), exhibit a phantom divide crossing. See the main text for the important implications of this fact. }
    \label{fig:Eos_and_rhoDE}
\end{figure*}

\begin{equation}\label{eq:DIC}
   {\rm DIC} =  \chi^2(\bar{\theta})+2p_D\,,
\end{equation}
with $p_D=\overline{\chi^2}-\chi^2(\bar{\theta})$ the effective number of parameters in the model, $\overline{\chi^2}$ the mean value of $\chi^2$ and $\bar{\theta}$ the mean of the parameters entering the Monte Carlo analysis. In contrast to AIC, DIC takes into account the shape of the full posterior distribution. However, when deviations from Gaussianity in the latter are important DIC can yield unrealistic values and this is why for some models we opt not to display their DIC values in the tables. We define the differences $\Delta {\rm AIC} \equiv {\rm AIC}_{\Lambda{\rm CDM}}-{\rm AIC}_j$ between the $\Lambda$CDM AIC value and that of a model $j$, and the analogous DIC difference,  $\Delta {\rm DIC} \equiv {\rm DIC}_{\Lambda{\rm CDM}}-{\rm DIC}_j$. A positive difference in these information criteria indicates a better performance of the dynamical DE models than the $\Lambda$CDM. According to Jeffreys' scale, if $0 \leq \Delta\textrm{DIC}<2$ one finds \textit{weak evidence} in favor of the model $j$, compared to the standard model. If $2 \leq \Delta\textrm{DIC} < 6$, we speak instead of \textit{positive evidence}. If $6 \leq \Delta\textrm{DIC} < 10$, there is \textit{strong evidence} in favor of the dynamical DE models, whilst if  $\Delta\textrm{DIC}>10$ we can conclude that there is \textit{very strong evidence} supporting the new model against the $\CC$CDM. An analogous consideration can be made using AIC, of course. We will show that the values of $\Delta\textrm{AIC}$ and $\Delta\textrm{DIC}$ that we obtain are fully consistent.

\begin{table*}[t!]
\centering
\begin{tabular}{|c ||c |c | c | c| c|  }
 \multicolumn{1}{c}{} & \multicolumn{1}{c}{} & \multicolumn{1}{c}{} & \multicolumn{1}{c}{} & \multicolumn{1}{c}{} & \multicolumn{1}{c}{}
\\\hline
{\small Parameter} & {\small $\Lambda$CDM} & {\small CPL} & {\small $Q_{\rm dm}$} & {\small $Q_{\Lambda}$} & {$w$XCDM}
\\\hline
$10^2\omega_b$ & $2.229\pm 0.013$ & $2.221\pm 0.013$ &  $2.214\pm 0.015$ & $2.234\pm 0.0013$ & $2.225^{+0.013}_{-0.014}$\\\hline
$10\,\omega_{\rm dm}$ & $1.180\pm 0.006$ & $1.191\pm 0.008$ & $1.178\pm 0.006$ & $1.283^{+0.063}_{-0.056}$  & $1.187\pm 0.010$ \\\hline
$\ln(10^{10}A_s)$ & $3.046\pm 0.014$ & $3.036\pm 0.014$ & $3.040\pm 0.014$ & $3.050^{+0.014}_{-0.015}$ & $3.041\pm 0.014$ \\\hline
$n_s$ & $0.967\pm 0.003$ & $0.965\pm 0.004$ & $0.964\pm 0.004$ & $0.969\pm 0.004$ & $0.966\pm 0.003$\\\hline
$\tau$ & $0.058\pm 0.007$ & $0.053\pm 0.007$ & $0.054\pm 0.007$ & $0.061\pm 0.008$ & $0.055\pm 0.007$ \\\hline
$H_0$ [km/s/Mpc] & $68.03\pm 0.29$ & $66.73\pm 0.56$& $68.54\pm 0.40$ & $67.24\pm 0.53$ & $67.23^{+0.53}_{-0.54}$\\\hline
$\nu$ & $-$ & $-$ & $-0.00063\pm 0.00034$ & $0.036\pm 0.021$ & $-$\\\hline
$w_0$ & $-$ & $-0.757\pm 0.057$ & $-$ & $-$ & $-$ \\\hline
$w_a$ & $-$ & $-0.83^{+0.23}_{-0.21}$  & $-$ & $-$ & $-$\\\hline
$w_X$ & $-$ & $-$ & $-$ & $-$  & uncons. \\\hline
$w_Y$ & $-$ & $-$ & $-$ & $-$  & $-0.942\pm 0.024$ \\\hline
$z_t$ & $-$ & $-$ & $-$ & $-$ & $3.15^{+0.01}_{-0.80}$ \\\hline\hline
$M$ [mag] & $-19.3932\pm 0.0080$ & $-19.386\pm 0.011$ & $-19.3932\pm 0.0080$  & $-19.386\pm 0.011$ & $-19.397\pm 0.011$ \\\hline
$\Omega_m$ & $0.305\pm 0.004$ & $0.319\pm 0.006$& $0.299\pm 0.005$ & $0.335\pm 0.018$ & $0.313\pm 0.005$\\\hline
$\sigma_{12}$ & $0.806\pm 0.006$ & $0.802\pm 0.007$& $0.805\pm 0.008$  & $0.745^{+0.024}_{-0.031}$ & $0.794^{+0.006}_{-0.007}$\\\hline
$S_{8}$ & $0.801\pm 0.009$ & $0.827\pm 0.009$ & $0.820\pm 0.009$ & $0.791^{+0.013}_{-0.015}$ & $0.816\pm 0.008$\\\hline\hline
$\chi^2_{\rm min}$ & $12636.02$ & $12616.82 $& $12633.42$ & $12632.98$ & $12624.67$\\\hline
$\Delta \text{AIC}$ & $-$ & $+15.20$ & $+0.6$ & $+1.04$ & $+5.35$\\\hline
$\Delta \text{DIC}$ & $-$ & $+16.64$ & $+1.64$ & $+1.16$ & $-$\\\hline
$E_{\Lambda {\rm CDM}}$ &  $-$ & $3.98\sigma$ & $1.61\sigma$ & $1.74\sigma$ & $-$  
\\\hline
\end{tabular}                
\caption{As in Table \ref{tab:table1}, but using CMB+DES-Y5+DESI.}\label{tab:table3}    
\end{table*}

As mentioned above, we reinforce our analysis and the conclusions extracted from it by estimating the exclusion level of the $\Lambda$CDM versus all the dynamical DE models presented in the previous section -- except for those that are not nested to the standard model (such as the $w$XCDM) -- with the precious help of the likelihood-ratio test, recently employed in, e.g., \cite{DESI:2025zgx,Gonzalez-Fuentes:2025lei}. Because of Wilks' theorem \cite{Wilk1938}, the distribution of the difference between the values of $\chi^2_{\rm min}$ obtained in a given model $M_A$ and a nested one $M_B$, i.e., $\Delta\chi^2_{\rm min}=\chi^2_{{\rm min},A}-\chi^2_{{\rm min },B}$, is a $\chi^2_{\alpha}$ distribution with $\alpha=N_B-N_A$ degrees of freedom, with $N_B>N_A$. Therefore, using that $\chi^2_\alpha$ distribution and the  value of $\Delta\chi^2_{\rm min}$ obtained in our fitting analysis, we can compute the associated $p$-value, i.e., the confidence level at which we can reject the null hypothesis, which in this case is the validity of model $M_A$, the $\Lambda$CDM. Then, in the last step, we translate these $p$-values in terms of the number of sigmas, $\xi$,  solving the following equation, 

\begin{equation}\label{eq:pvalue}
p{\rm -value} = 1-\frac{1}{\sqrt{2\pi}}\int_{-\xi}^{\xi}e^{-y^2/2}dy\,,
\end{equation}
as if the $p$-value came from a univariate Gaussian. We define the exclusion level of $\Lambda$CDM simply as $E_{\Lambda{\rm CDM}}=\xi\times\sigma$. Again, the values of $E_{\Lambda{\rm CDM}}$ obtained for the various dynamical DE models are shown in the tables. We find that they resonate extremely well with the corresponding values of the information criteria in our analysis.

\textcolor{black}{Before concluding this section, we want to remark that, in this work, we avoid the computation of exact Bayes factors \cite{Trotta:2008qt}, since they are highly dependent on the choice of priors, even if the latter are uninformative and do not affect the shape of the posterior distributions, see, e.g., \cite{Efstathiou:2008ed}. We just want to remark here that computing the exact Bayes ratios alone in the absence of solid and physically motivated priors is not enough to extract
solid conclusions; in these cases, one has to show what is the dependence of the Bayes factors on the prior volume
(see, e.g., \cite{SolaPeracaula:2018wwm,Patel:2024odo}), or perform frequentist-Bayesian
studies as those proposed in  \cite{Keeley:2022ojz,Amendola:2024prl}, which can help to remove the
unwanted effect of uninformative priors, but imply in general a huge computational cost, especially in studies that analyze many models and with several datasets, as it is the case in this work.}


\section{Results}\label{sec:results}

\begin{figure*}[t!]
    \centering
    \includegraphics[width=0.49\linewidth]{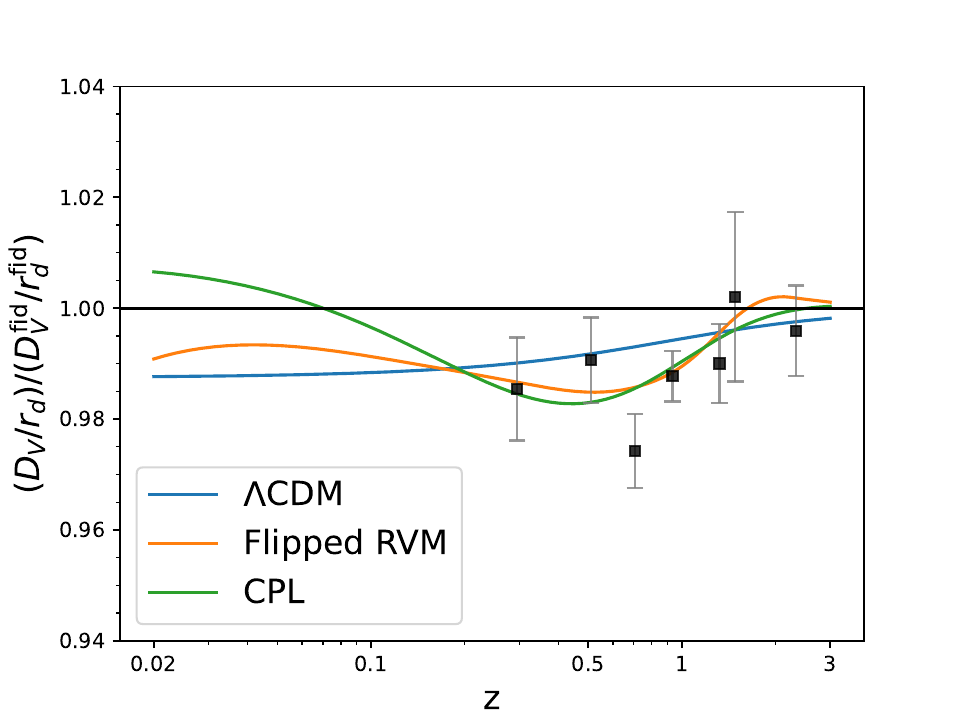}
    \hfill
    \includegraphics[width=0.49\linewidth]{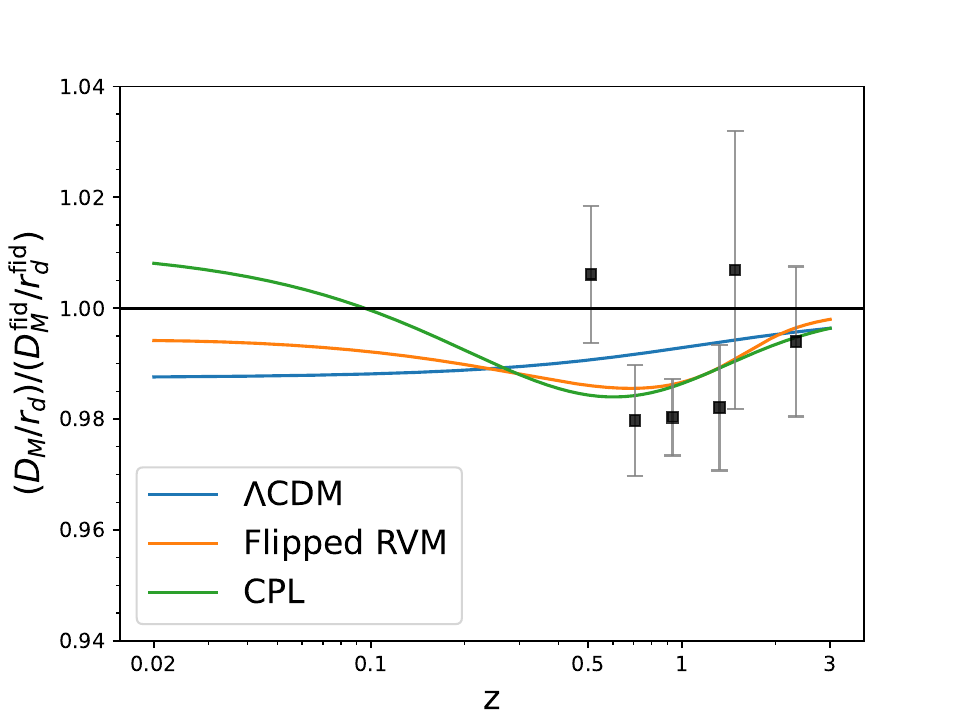}
    
    \vspace{0.2cm} 

    \includegraphics[width=0.49\linewidth]{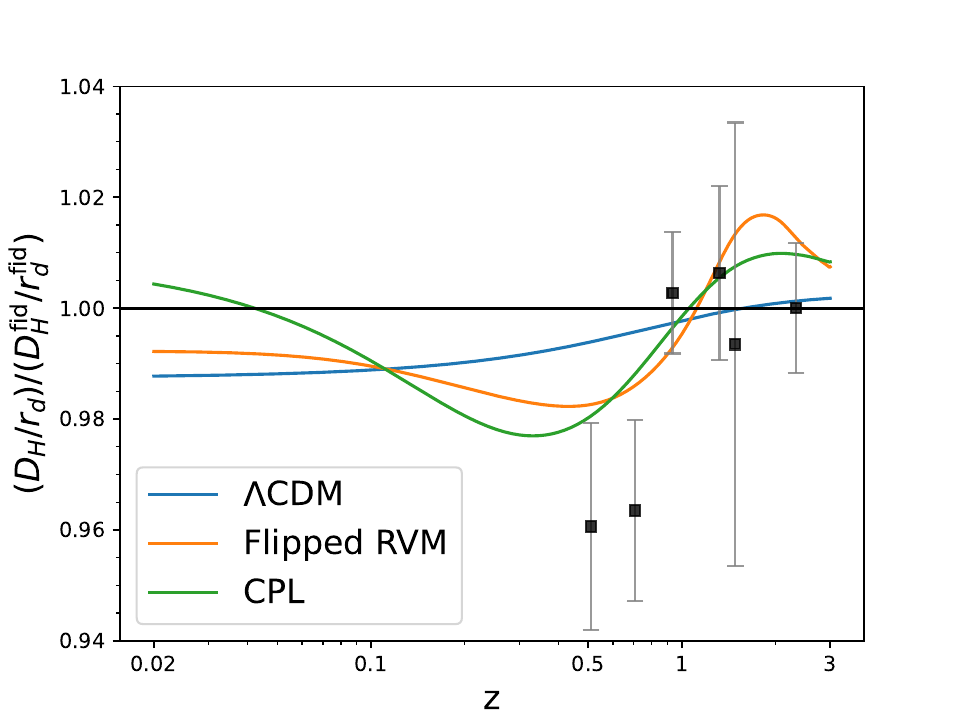}
    \hfill
    \includegraphics[width=0.49\linewidth]{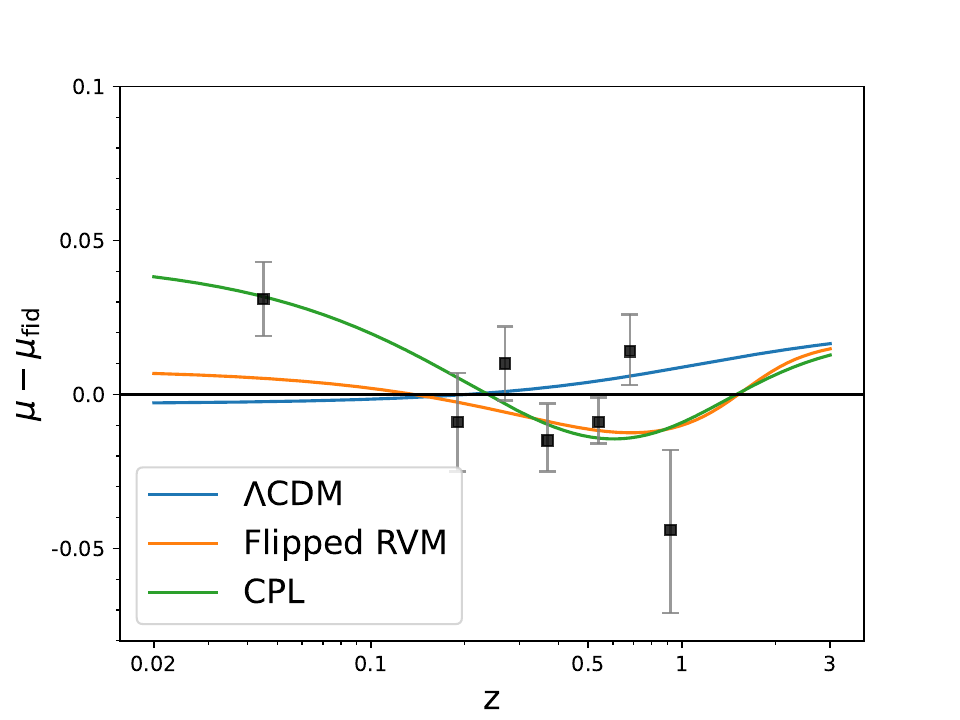}
    
    \caption{\textcolor{black}{Residual plots for the best-fit $\Lambda$CDM, Flipped RVM and CPL models obtained with the dataset CMB+DES-Y5+DESI. We normalize them with respect to the Planck 2018 TT,TE,EE+lowE+lensing $\Lambda$CDM results \cite{Planck:2018vyg}. In the top left, top right and bottom left panels, we have the curves for the observables employed by DESI, together with the BAO DESI DR2 \cite{DESI:2025zgx} data points and their corresponding error bars. In the bottom right panel we have the normalized curves for the distance moduli; as done in \cite{DESI:2025zgx},  for the SNIa data from DES-Y5 \cite{DES:2024hip,DES:2024jxu}, we show binned data points to improve the visualization of the theoretical curves of the various models. }}
    \label{fig:residual_plots}
\end{figure*}

\begin{figure*}[t!]
    \centering    \includegraphics[width=0.8\linewidth]{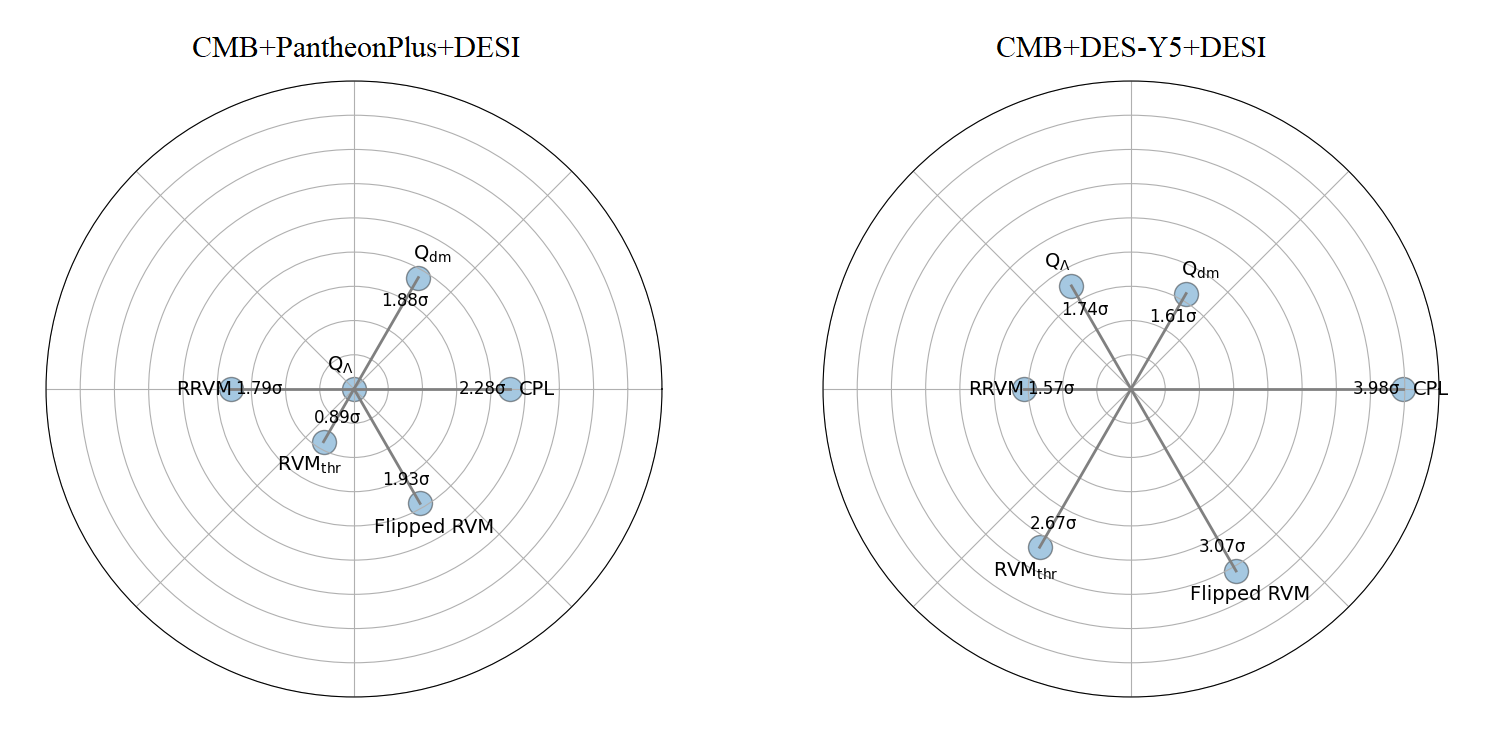}
    \caption{Exclusion level $E_{\Lambda{\rm CDM}}=\xi\times\sigma$ -- see Eq.\,\eqref{eq:pvalue} --  of the standard model $\Lambda$CDM with respect to the dynamical DE models that are nested to the former, obtained from the fitting analyses with CMB+PantheonPlus+DESI (left plot) and CMB+DES-Y5+DESI (right plot).}
    \label{fig:E_comparison}
\end{figure*}

\begin{figure*}[t!]
    \centering    \includegraphics[width=0.8\linewidth]{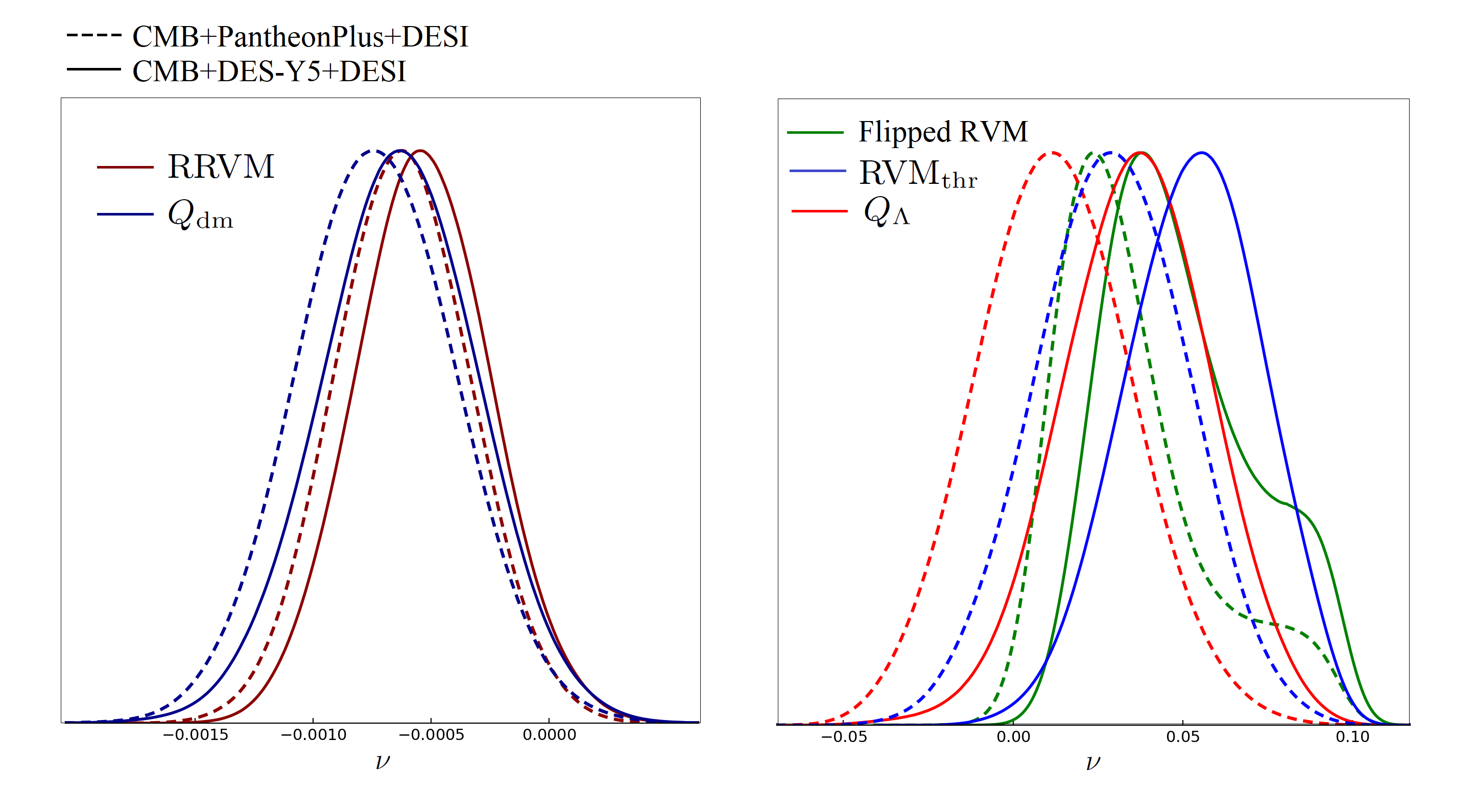}
    \caption{Posterior distributions of the parameter $\nu_i$ that controls the transfer of energy between the vacuum and dark matter in the various interacting models (cf. Sec. \ref{sec:models}), obtained with the datasets CMB+PantheonPlus+DESI (dashed curves) and CMB+DES-Y5+DESI (solid curves). In the figure, abscissas denotes the parameter $\nu$ of the RVM$_{\rm thr}$ and RRVM,  but also  $\nu_\Lambda$ and $\nu_{\rm dm}$ of $Q_\Lambda$ and $Q_{\rm dm}$, respectively.}
    \label{fig:posteriors_nu}
\end{figure*}

Our results are displayed in the fitting Tables \ref{tab:table1}-\ref{tab:table4} and in Figs. \ref{fig:Eos_and_rhoDE}-\ref{fig:posteriors_nu}. We discuss the results obtained with the two data sets employed in our analysis in turn.
\newline
\newline
\noindent {\it (i) Results with dataset CMB+PantheonPlus+DESI}
\newline
\newline
Let us discuss first the results obtained with CMB+PantheonPlus+DESI -- our most conservative dataset --, which are collected in Tables \ref{tab:table1} and \ref{tab:table2}. The two models that provide the best description of this dataset are the CPL (i.e. the $w_0w_a$CDM parameterization) and the flipped RVM. Both lead to a decrease of $\sim 7.5$ units of $\chi^2_{\rm min}$ compared to the $\Lambda$CDM. However, the flipped RVM has one more extra parameter than the CPL (3 vs 2), which results in a stronger penalization from the perspective of Occam's razor. The $\Lambda$CDM is excluded at $\sim 2.3\sigma$ CL within the CPL, and at $1.9\sigma$ CL within the flipped RVM; the information criteria yield a consistent result, namely $\Delta$AIC and $\Delta$DIC point to a positive evidence for the CPL compared to the standard model, while for the flipped RVM the value of $\Delta$AIC is not conclusive, yielding only a weak evidence for it. For this model, given the presence of large non-Gaussian features in the posterior distributions, we prefer to avoid the use of $\Delta$DIC. The decrease of $\chi^2_{\rm min}$ in the two models can be easily understood by looking at the two upper plots of Fig. \ref{fig:Eos_and_rhoDE}. In particular, one can see that the best-fit effective dark energy density in these two models lies well inside the 68\% CL band of the model-agnostic reconstruction of Ref. \cite{Gonzalez-Fuentes:2025lei}, which is shown in light blue in that figure. Although there are some important differences between them, both models are capable of explaining a transition from small values of the effective (self-conserved) DE density  $\tilde{\rho}_{\rm de}(z)$ (Eq. \ref{eq:effectiveDens}) at $z\sim 2.5$, followed by a peak at $z\sim 0.4-0.8$ and a subsequent decrease of $\tilde{\rho}_{\rm de}(z)$ to similar values at present.

\begin{table*}[t!]
\centering
\begin{tabular}{|c ||c |c | c | c|   }
 \multicolumn{1}{c}{} & \multicolumn{1}{c}{} & \multicolumn{1}{c}{} & \multicolumn{1}{c}{} & \multicolumn{1}{c}{}
\\\hline
{\small Parameter} & {\small $\Lambda$CDM}  & {\small Flipped RVM} & {\small RRVM} & {\small RVM$_{\text{thr}}$}
\\\hline
$10^2\omega_b$ &  $2.229\pm 0.013$  & $2.216^{+0.013}_{-0.015}$ & $2.214\pm 0.015$  & $2.235\pm 0.013$ \\\hline
$10\,\omega_{\rm dm}$ & $1.180\pm 0.006$ & $1.340\pm 0.058$ & $1.178\pm 0.006$  & $1.337\pm 0.061$\\\hline
$\ln(10^{10}A_s)$ & $3.046\pm 0.014$  & $3.040\pm 0.014$ & $3.040\pm 0.014$ & $3.051 \pm 0.015$\\\hline
$n_s$ & $0.967\pm 0.003$  & $0.963\pm 0.004$ & $0.964\pm 0.004$  & $0.969\pm 0.004$ \\\hline
$\tau$ & $0.058\pm 0.007$  & $0.054^{+0.007}_{-0.008}$ & $0.054\pm 0.007$ & $0.061\pm 0.007$ \\\hline
$H_0$ [km/s/Mpc] & $68.03\pm 0.29$   & $67.46\pm 0.53$ & $68.52\pm 0.41$  & $66.82\pm 0.54$ \\\hline
$\nu$ & $-$ & $0.046^{+0.013}_{-0.023}$ & $-0.00053\pm 0.00029$  & $0.053\pm 0.020$ \\\hline
$z_1$ &  $-$  & $2.56^{+0.77}_{-0.90}$ & $-$ &  $-$ \\\hline
$z_2$ & $-$ & $5.3^{+1.6}_{-2.9}$ & $-$ &$-$ \\\hline\hline
$M$ [mag] & $-19.3932\pm 0.0080$ & $-19.391\pm 0.011$ & $-19.380\pm 0.011$ &  $-19.409\pm 0.010$ \\\hline
$\Omega_m$ & $0.3045\pm 0.0037$  & $0.345\pm 0.017$ & $0.2995 \pm 0.0046$ & $0.351\pm 0.019$ \\\hline
$\sigma_{12}$ & $0.806\pm 0.006$ & $0.734\pm 0.023$ & $0.805\pm 0.008$ & $0.725^{+0.024}_{-0.028}$ \\\hline
$S_{8}$ & $0.801\pm 0.009$  & $0.793^{+0.013}_{-0.011}$ & $0.820\pm 0.009$  & $0.784 \pm 0.013$ \\\hline\hline
$\chi^2_{\rm min}$ & $12636.02$ & $12621.36$ & $12633.54$  & $12628.90$  \\\hline
$\Delta\text{AIC}$ & $-$ &$+8.66$ & $+0.48$ & $+5.12$ \\\hline
$\Delta\text{DIC}$ & $-$ &$-$ & $+1.78$ &  $+4.64$ \\\hline
$E_{\Lambda {\rm CDM}}$ &  - & $+3.07\sigma$ &1.57$\sigma$   & $2.67\sigma$
\\\hline
\end{tabular}                        
\caption{As in Table \ref{tab:table2}, but using CMB+DES-Y5+DESI.}\label{tab:table4}  
\end{table*}

These models yield values of the Hubble parameter in the low range, at odds with the local distance measurements of SH0ES \cite{Riess:2021jrx,Riess:2024vfa}. This is of course expected, since it is well-known that late-time modifications of the expansion history alone cannot induce an important raise of $H_0$ without spoiling the good fit to the anisotropic BAO and CMB data \cite{Sola:2017znb,Knox:2019rjx,Krishnan:2021dyb,Lee:2022cyh,Keeley:2022ojz,Gomez-Valent:2023uof,Pedrotti:2025ccw}. This applies to all the models analyzed in this paper, which can only reach values of $H_0$ in the ballpark of $67-69$ km/s/Mpc, and values of the absolute magnitude of SNIa close to $M=-19.40$ mag.

Regarding the structure formation, there are indeed substantial differences between the predictions of the various dynamical DE models. Although in this work we have not employed LSS data in our fitting analyses for the reasons explained in Sec. \ref{sec:data}, we deem it important to comment on the predictions of the amplitude of matter fluctuations in each of these models, since they can play a non-negligible role when growth data are considered, and also in the discussion of the growth tension. This is why in our tables we display the values of the parameters $\sigma_{12}$ -- i.e., the rms mass fluctuations at a fixed scale of 12 Mpc -- and $S_8=\sigma_8(\Omega_m/0.3)^{0.5}$. Quantifying the amplitude of the matter power spectrum with $\sigma_{12}$ instead of $\sigma_8$ turns out to be a better option, since $\sigma_8$ is evaluated at a scale $R_8=8/h$ Mpc, which depends on the value of $H_0$, whereas $\sigma_{12}$ is a fixed value independent of $h$. This fact might introduce a non-negligible bias in the case of $\sigma_8$ when comparing the performance of models with different values of the Hubble parameter, simply because in this case we would compare characteristic amplitudes of matter fluctuations at different scales \cite{Sanchez:2020vvb,Forconi:2025cwp}. For this reason the parameter $\sigma_{12}$ is claimed to be more suited for clustering analyses, see, e.g. \cite{Semenaite:2021zxw,Semenaite:2022unt}. On the other hand, $S_8$ is the observable inferred in weak lensing studies \cite{Secco:2022kqg,Garcia-Garcia:2024gzy}. We find that the values of these two parameters obtained in the CPL and the flipped RVM are compatible at $1\sigma$ CL with those obtained in the $\Lambda$CDM. However, in the CPL the central values lie above those of the standard model, while in the flipped RVM they lie below. In fact, in this model the central value of $\sigma_{12}\sim 0.76$ is pretty small. This is an interesting feature of the model which could be exploited if in the future LSS data favor a less clumpy universe than in the $\Lambda$CDM. This suppression of growth can be understood with the aid of Eq. \eqref{eq:pert} as follows: for a positive $\nu$ as the one preferred by the data,  at $z<z_1$ (with $z_1=1.22$ for the best fit), $\rho_{\rm vac}^\prime<0$ and therefore $\delta_{\rm dm}^\prime$ receives a negative contribution because of the interaction, which is non-negligible since $\nu\sim \mathcal{O}(10^{-2})$. This fights, of course, against the aggregation of matter. At $z_1<z<z_2$ (with $z_2=2.16$ for the best fit), the sign of $\nu$ is flipped and therefore it has the opposite effect. However, we have explicitly checked numerically (using the best-fit values) that the matter energy fraction $\Omega_m(a)$ is smaller than in $\Lambda$CDM in a significant portion of that redshift range, which somehow softens the clustering effects, allowing for an overall decrease of matter fluctuations in the late universe. The weak-lensing measurements from the Dark Energy Survey (DES),
$S_8 = 0.775^{+0.026}_{-0.024}$ \cite{DES:2021wwk},
the Hyper Suprime-Cam (HSC),
$S_8 = 0.763^{+0.040}_{-0.036}$ \cite{Miyatake:2023njf},
and the Kilo-Degree Survey (KiDS),
$S_8 = 0.815^{+0.016}_{-0.021}$ \cite{Wright:2025xka},
are all in perfect agreement -- at the $\sim 1\sigma$ confidence level -- 
with the values inferred in $\Lambda$CDM and the flipped RVM, for which we obtain
$S_8 = 0.810 \pm 0.008$ and $S_8 = 0.800^{+0.012}_{-0.010}$, respectively.
For the CPL model, instead, we find $S_8 = 0.822 \pm 0.009$, which is again slightly
larger than in the standard model and the flipped RVM.
Although this value is fully compatible with KiDS, it exhibits a mild tension
with the DES measurement at the $\sim 1.8\sigma$ level, which decreases to
$\sim 1.5\sigma$ in the case of HSC.

For the $Q_{\rm dm}$ and RRVM models, the level of exclusion of $\Lambda$CDM is close to $1.9\sigma$ CL, similar to that obtained for the flipped RVM, despite the former ($Q_{\rm dm}$ and RRVM) having $\chi^2_{\rm min}$ values larger by about $4-4.5$ units than the flipped RVM. The reason is that these models introduce only one additional parameter relative to the standard model, instead of three, and therefore their complexity is penalized less. The best-fit curves of the effective EoS parameter and the dark energy density for these models lie in the $68\%$ CL band at low redshifts (cf. again the upper plots of Fig. \ref{fig:Eos_and_rhoDE}), but fall outside it for $z\gtrsim1.2$, although they are still in the 95\% CL region. The constraints on $\nu$ and $\nu_{\rm dm}$ in the RRVM and $Q_{\rm dm}$ are extremely similar. This is a clear manifestation of the resemblance of the source functions in both models, which in turn also lead to pretty similar expressions for the DM and vacuum energy densities (cf. Eqs. \ref{eq:rhoDMRRVM}-\ref{eq:rhoLRRVM} and Eqs. \ref{eq:rhodmQdm}-\ref{eq:rhovacQdm}). These models do not exhibit an effective crossing of the phantom divide. The effective EoS remains in the phantom region $\forall{z}$ and approaches $w_{\rm eff, de}(z)$ from below when $z\to 0^+$. Current data constrain $|\nu|$ below $\mathcal{O}(10^{-3})$, and prefer negative values at $2\sigma$ CL. The negative sign of $\nu_{\rm dm}$ and $\nu$ for these models is important in order to understand why we obtain for them slightly larger values of $\sigma_{12}$ and $S_8$ than in the $\Lambda$CDM (cf. again Eq. \ref{eq:pert}). They are, though, still compatible at $1\sigma$ CL. The inclusion of redshift-space distortion or weak lensing data in the fitting analysis favoring a lesser amount of structure than in the standard model would shift the values of $\nu_{\rm dm}$ and $\nu$ toward the region of less negative values or even to the region of positive values -- see, e.g., \cite{Gomez-Valent:2017idt,Gomez-Valent:2018nib} for a comprehensive discussion of density fluctuations in the RVM.

As for the $Q_{\Lambda}$ and RVM$_{\rm thr}$ models, they do not significantly improve the description of the data relative to  $\Lambda$CDM for the current dataset, despite having one additional parameter. According to the information criteria, these models are either only weakly favored or weakly disfavored with respect to the standard model. However, it is interesting to note that they can exhibit a crossing of the phantom divide, even though this crossing occurs at somewhat larger redshifts, close to $z\sim 1$, and the effective DE density is outside the 68\% CL band of the reconstructed function from \cite{Gonzalez-Fuentes:2025lei} in almost all the redshift range covered in Fig. \ref{fig:Eos_and_rhoDE}. The small improvement of the $Q_{\rm dm}$ and RRVM models with respect to $Q_\Lambda$ and RVM$_{\rm thr}$ is essentially due to the better behavior of $\tilde{\rho}_{\rm de}(a)$ at $z<0.5$; in the latter models, this function remains in the $1\sigma$ region of the reconstructed bands, while it is outside that region in the former models. The values of $\sigma_{12}$  are considerably smaller than in the standard model, with central values $\sigma_{12}\sim 0.76$ in the RVM$_{\rm thr}$ case and $\sigma_{12}\sim 0.78$ in the  $Q_\Lambda$ model. This is due to the relatively large positive values of  $\nu_{\rm dm}$ and $\nu$ allowed by the data, which leave a similar imprint on large-scale structure to that discussed above for the flipped RVM at $z<z_1$.

Finally, we note that the $w$XCDM model (and a fortiori the $\Lambda_s$CDM model, being a particular case) is not preferred over the concordance model in the light of CMB+PantheonPlus+DESI. In spite of being able to decrease by 4.4 units the value of $\chi^2_{\rm min}$ compared to the $\Lambda$CDM, the $\Delta$AIC criterion clearly tells us that this improvement is not sufficient to select the $w$XCDM, since it receives a severe penalization due to the use of three additional parameters. The presence of strong non-Gaussian features in this model prevents us from displaying the value of 
$\Delta$DIC in this case; in addition, we do not compute $E_{\Lambda{\rm CDM}}$ for this model, since it is not nested to the $\Lambda$CDM. The preferred EoS of dark energy at low redshifts is of quintessence type, $w_Y=-0.970^{+0.024}_{-0.025}$ at 68\% CL, but is still compatible with a cosmological constant at $\sim 1.2\sigma$ CL. Before the transition, at $z>z_t$, the EoS $w_X$ of the $X$ component -- recall, $X$ has negative energy density and positive pressure -- is not well constrained by this dataset and, indeed, its posterior is fully dominated by the prior (cf. Sec. \ref{sec:PM}). The transition redshift lies in the region $z_t\in[2.3,3.6]$ at 68\% CL. Both, the values of $H_0$ and $\sigma_{12}$ are compatible with those of the standard model.
\newline\newline
\noindent \noindent {\it (ii) Results with dataset CMB+DES-Y5+DESI}
\newline
\newline
We now turn our attention to the results based on the CMB+DES-Y5+DESI dataset, obtained by replacing the Pantheon+ SNIa with the DES-Y5 sample. They are reported in Tables \ref{tab:table3}-\ref{tab:table4}, \textcolor{black}{and are also reinforced by Table \ref{tab:tableAppendix}}. We find an overall enhancement in the evidence for DE dynamics among all models capable of accommodating an effective crossing of the phantom divide when confronted with the aforementioned data. This applies to all the DE models considered here, except for  $Q_{\rm dm}$ and RRVM (without threshold).  The mentioned enhancement for the models exhibiting crossing is to be expected, since the DES-Y5 SNIa sample shows a preference for this feature. This class includes in our case the CPL and flipped RVM models, as well as the $Q_\Lambda$ and RVM$_{\rm thr}$ scenarios. In particular, the exclusion level $E_{\Lambda{\rm CDM}}$ of $\Lambda$CDM relative to the CPL increases to approximately $\sim 4\sigma$ CL, with the information criteria indicating very strong evidence. For the flipped RVM, the exclusion reaches $\sim 3.1\sigma$ CL, with a $\Delta$AIC value lying in the strong-evidence regime. \textcolor{black}{The CPL and flipped RVM emerge as the models providing the best fits to the data. The origin of this
improvement can be understood by examining the breakdown of the $\chi^2_{\rm min}$ contributions shown
in Table~\ref{tab:tableAppendix}. While both models reduce $\chi^2_{\rm CMB}$ by only $2\!-\!3$ units,
the dominant gain relative to $\Lambda$CDM arises from their improved description of the BAO and SNIa
datasets. In particular, $\chi^2_{\rm DESI}$ decreases by 10 and 8 units, and $\chi^2_{\rm DES-Y5}$ by
7 and 6 units, for the CPL and flipped RVM models, respectively. This is of course perfectly consistent with the corresponding residual plots, cf. Fig. \ref{fig:residual_plots}.}

\begin{table}[t!]
\centering
\renewcommand{\arraystretch}{1.8}
\begin{tabular}{|c|c|c|c|c|}
\hline
Model & $\chi^2_{\rm CMB}$ & $\chi^2_{\rm DES-Y5}$ & $\chi^2_{\rm DESI}$ & $\chi^2_{\text{min}}$ \\ \hline
$\Lambda\text{CDM}$ &   10973.03    &   1648.05     &    14.94 &  12636.02 \\ \hline
$\text{CPL}$ &  10970.37      &  1638.17     & 8.30     & 12616.82  \\ \hline
$Q_{\text{dm}}$ &  10971.36      & 1650.30      &  11.77    & 12633.43  \\ \hline
$Q_\Lambda$ & 10977.23  &  1642.85  &  12.91 & 12632.99 \\ \hline
$w\text{XCDM}$      &   10972.08    &   1639.88    &   12.71   &  12624.67 \\ \hline
$\text{Flipped} \text{RVM}$ & 10971.91 &  1640.53 &  8.93 & 12621.37       \\ \hline
$\text{RRVM}$ & 10971.53 &  1651.31&  10.70 &  12633.54 \\ \hline
$\text{RVM}_{\text{thr}}$& 10973.00 & 1640.27 & 15.63  & 12628.90       \\ \hline
\end{tabular}
\caption{\textcolor{black}{Individual $\chi^2_i$ contributing to $\chi^2_{\rm min}$, obtained in the fitting analyses for the various models with
CMB+DES-Y5+DESI, corresponding to the results displayed in Tables \ref{tab:table3} and \ref{tab:table4}.}}\label{tab:tableAppendix} 
\end{table}

In the case of the $Q_\Lambda$ and RVM$_{\rm thr}$ models, we find -- again, in terms of $E_{\Lambda{\rm CDM}}$ -- preferences at  $\sim 1.7\sigma$ CL and $\sim 2.7\sigma$ CL, respectively, relative to $\Lambda$CDM. As discussed above, the $Q_{\rm dm}$ and RRVM models do not exhibit a phantom-divide crossing for the best-fit values obtained in our analysis, and consequently their statistical preference remains essentially unchanged, at the level of $\sim 1.6\sigma$ CL. Under the present dataset, and in contradistinction to the previous one,  the model $w$XCDM offers a pretty good fit to the data and is strongly preferred over the $\Lambda$CDM, according to its $\Delta$AIC value. However, the performance of this model is somehow mitigated in the presence of anisotropic BAO data and without the SH0ES prior. This is in contrast to the improved results obtained when the SH0ES prior is used  with  anisotropic and, specially,  with  angular BAO data, as demonstrated in  Ref. \cite{Gomez-Valent:2024ejh}.  In Fig. \ref{fig:E_comparison}, we compare in a visual way the values of $E_{\Lambda{\rm CDM}}$ (cf. Sec. \ref{sec:data}, and Eq. \ref{eq:pvalue}) obtained for all the nested models to the $\Lambda$CDM that have been analyzed in this paper. Recall that the $w$XCDM does not appear in this figure since it is not nested to the $\CC$CDM. The figure eases, we hope,  the comparison between the results obtained considering Pantheon+ and DES-Y5.

The constraints in all models with the current dataset remain consistent with those obtained using Pantheon+, but the parameters governing the DE dynamics are generally shifted further away from the values corresponding to the standard model, except for the $Q_{\rm dm}$ and RRVM, since we have already mentioned that the evidence for these models is a bit smaller with DES-Y5, due to the absence of the crossing feature that is favored by DES-Y5 data. In Fig. \ref{fig:posteriors_nu}, we show the one-dimensional posterior distributions for the parameters $\nu$, $\nu_\Lambda$ and $\nu_{\rm dm}$ of the running vacuum models, $Q_\Lambda$ and $Q_{\rm dm}$, respectively.

One aspect worth emphasizing is that, in the $Q_\Lambda$, RVM$_{\rm thr}$, and flipped RVM models, we find a significantly more suppressed growth than that found with Pantheon+\footnote{The strong suppression of growth found in the RVM$_{\rm thr}$ was already emphasized in our previous works \cite{SolaPeracaula:2021gxi,SolaPeracaula:2023swx}.}. When DES-Y5 data are used, the values of $\sigma_{12}$ lie approximately $2\sigma$ below the $\Lambda$CDM prediction in the $Q_\Lambda$ case, and about $3\sigma$ below it in the RVM$_{\rm thr}$ and flipped RVM models. The effect is less pronounced in the weak-lensing parameter; nevertheless, we still observe a mild suppression in terms of $S_8$. A similar level of tension  (of about $2-3\sigma$) both in $\sigma_{12}$ and $S_8$ is found between the values of these parameters in the aforesaid dynamical DE models and the CPL prediction. Hence, LSS observations could eventually be used to discriminate between these models. At present, weak-lensing measurements are unable to efficiently do so among the various models, partly because the central values reported by different surveys span a relatively broad range. In the CPL parameterization, we find $S_8=0.827\pm 0.009$, so the tension with the DES weak-lensing measurement increases to the $\sim 2\sigma$ CL, while it remains milder for HSC and fully consistent with KiDS. In the $Q_\Lambda$, RVM$_{\rm thr}$, and flipped RVM models, the inferred values are $S_8=0.791^{+0.013}_{-0.015}, 0.784\pm 0.013$ and $0.793^{+0.013}_{-0.011}$, respectively. They are still compatible at the $\sim 1\sigma$ CL with the available DES, HSC, and KiDS data. Nevertheless, as already noted, future improvements in the precision of weak-lensing measurements could help to discriminate among the different dynamical dark energy models considered in this work.

Returning to Fig. \ref{fig:Eos_and_rhoDE}, we can see in the lower plots the best-fit curves of the effective DE density and EoS parameter for the various models under study, obtained from the fitting analysis with CMB+DES-Y5+DESI. They look very similar to those obtained with Pantheon+, but now some of the curves fall outside the $95\%$ CL band in some redshift ranges,  since the uncertainties of the reconstructions are smaller with DES-Y5. We can see that in this case only the CPL and the flipped RVM remain in most of the relevant region.

All in all, as expected, we find a significant enhancement of the dynamical dark energy signal when DES-Y5 data are included in the analysis, not only for the CPL parameterization but also for other models that consider either interactions within the dark sector ($Q_{\rm dm}$ and $Q_\CC$) or a sudden transition of the dark energy component to negative energy densities at high redshift ($w$XCDM). In the case of the RVM, the data favor either the presence of a cutoff in the redshift (which we have also called `threshold') around the onset of the vacuum-dominated stage of the universe, as in the RVM$_{\rm thr}$, or a reversal in the direction of the energy flux, as in the flipped RVM.

\section{Conclusions}\label{sec:conclusions}

In this work, we have revisited the potential phenomenological impact of a number of models of dark energy (DE) of different kinds in the light of the latest observations. Some of these models are well-known and have been extensively investigated in the past. In fact, they have granted the firsts hints that modern cosmological data might show a preference for an evolving DE rather than just a rigid cosmological constant $\CC$ for the entire cosmic history. Needless to say, the eventual confirmation of this result would be of paramount importance in the history of cosmology since Einstein introduced the $\CC$ term in 1917.  For example, already 10 years ago a series of systematic studies on the running vacuum model (RVM) \cite{Sola:2013gha,Sola:2015rra,SolaPeracaula:2022hpd} confronting a large set of cosmological data of all sorts (SNIa+BAO+H(z)+LSS+BBN+CMB) intriguingly pointed to the possibility that the vacuum energy density (VED) of the universe might be evolving with the expansion at a fairly large confidence level of $\sim (3-4)$$\sigma$\, \cite{Sola:2015wwa,Sola:2016jky}. The claim was further supported in subsequent works\,\cite{SolaPeracaula:2016qlq,SolaPeracaula:2017esw,Sola:2017znb} and is particularly encouraging since the RVM framework is not just a toy model or a parameterization; it is based on QFT in curved spacetime and hence its theoretical structure is accounted for from fundamental physics\,\cite{SolaPeracaula:2022hpd}.

Together with the canonical form of the RVM, and the RRVM variant of it, and including also the possibility of having a threshold in the VED behavior at $z_t=1$, we have revisited two traditional models of interactive vacuum energy, as well as a recently proposed composite model of DE, the $w$XCDM\,\cite{Gomez-Valent:2024tdb,Gomez-Valent:2024ejh}, in which the DE is made out of two dynamical components $(X,Y)$ acting in sequence in the expansion history. The $X$ component acts first in the past and  behaves as ``phantom matter'' \cite{Grande:2006nn} (a cosmic fluid possessing negative energy density but positive pressure), whereas the $Y$ component is more recent and behaves as conventional quintessence, this scenario being compatible with the DE analysis of the DESI Collab.\ \cite{DESI:2024mwx,DESI:2025zgx}. An interesting variant of the RVM, which we  have proposed and scrutinized here for the first time, is the ``flipped RVM'', in which the dynamical term of the VED flips sign in an intermediate redshift interval in our past, thereby enforcing a reversal in the direction of the energy flux between the interacting DE and  DM. This model bears relation with the $w$XCDM, in that it could also produce an effective phantom matter behavior. At the same time, we have fitted our datasets with the popular $w_0w_a$CDM (CPL) parameterization of the DE \cite{Chevallier:2000qy,Linder:2002et}, which we have used as a generic benchmark scenario; and,  of course,  we have compared all of these models and parameterizations with the standard $\CC$CDM model.

The models studied here have been confronted with the most recent observations.  The analysis was performed with the help of the standard Boltzmann code \texttt{CLASS} \cite{Lesgourgues:2011re,Blas:2011rf}. In this work, however, we have focused on just three standard sources of modern cosmological data: i) SNIa (comprising two datasets, one from Pantheon$+$ and the other  from DES-Y5, but treated separately); ii) BAO data from DESI DR2; and  iii) CMB data (from Planck PR4 TT+EE+TE+lensing measurements). The reason why we have restricted our analysis to these data sources only, is because we wish to ease as much as possible the comparison with other studies existing in the current literature. However, the impact of additional data sets (in particular the inclusion of growth data, which we have used on previous occasions; see the aforementioned references) cannot be underestimated and will be taken into account in future work.

Using CMB+PantheonPlus+DESI, we find mild hints of dynamical dark energy -- at the level of $\sim 2\sigma$ CL -- in several models with an interaction between dark matter and vacuum, close to those found with the CPL parameterization. Some of our models have special features that can be very relevant, even for future studies. For example, both the RVM$_{\rm thr}$ and the flipped RVM present the lowest fitted values of $\sigma_{12}$ and $S_8$ of the entire analysis.  In addition, the flipped RVM  leads to the lowest fitted value of $\chi^2_{\rm min}$ among all models when Pantheon+ is used, and comparable to that of the CPL with DES-Y5 data, although this model is penalized by the use of one additional parameter compared to the CPL. Remarkably, it can mimic quite efficiently an effective crossing of the phantom divide, in the line of those inferred in the CPL and model-agnostic reconstructions of the dark energy \cite{DESI:2024aqx,DESI:2025fii,Berti:2025phi,Gonzalez-Fuentes:2025lei}.

These hints grow significantly in some of the models studied in this paper when Pantheon+ is replaced with DES-Y5. For instance, in the RVM$_{\rm thr}$ and the flipped RVM the standard model is excluded at  $\sim 2.7\sigma$ and $\sim 3.1\sigma$ CL, which is however lower than the $\sim 4\sigma$ signal found with the CPL. Worth emphasizing, the evidence increases (above all, if not exclusively) in those models that are capable of explaining a crossing of the phantom divide. We find that the CPL and the flipped RVM are the only two models in our study which are able to remain within the 68\% and 95\% CL region of the model-agnostic reconstructions of Ref. \cite{Gonzalez-Fuentes:2025lei}, respectively.

To better assess these hints of DE dynamics, it will be instrumental to monitor the evolution of the SNIa data, since they are playing a crucial role in the enhancement of the signal. The recent re-calibration of DES-Y5 SNIa \cite{DES:2025sig} seems to point to downgrading of the evidence, which lies now closer to (but still higher than) that found with Pantheon+. It will be equally important to follow up on the status of the Hubble tension, since the models analyzed in this paper yield low estimates of $H_0$, in tension with SH0ES. This is, however, a feature of using BAO 3D, which is usually in conflict with the $H_0$ tension in the context of late-time dynamical DE or modified gravity models \cite{Sola:2017znb,Knox:2019rjx,Krishnan:2021dyb,Lee:2022cyh,Keeley:2022ojz,Gomez-Valent:2023uof,Pedrotti:2025ccw}.

We believe that our work contributes to stress  the following intriguing possibility: if the effective dark energy density were to exhibit a peak at intermediate redshifts, this would point to the presence of nontrivial physics beyond $\Lambda$CDM, including the possibility of interactions between dark matter and vacuum, and even changes in the direction of their energy flux (decay of DM to vacuum, or the other way around) throughout the cosmic history.


\section*{Acknowledgments}
JdCP's research was financially supported by the projects "Plan Complementario de I+D+i en el \'area de Astrof{\'\i}sica" and "Desarrollo de algoritmos de big data y data science aplicados a la física de partículas" funded by the European Union within the framework of the Recovery, Transformation and Resilience Plan - NextGenerationEU and by the Regional Government of Andaluc{\'i}a (References AST22\_00001 and AST22\_8.4\_SR). Also, JdCP acknowledges partial support from MICINN (Spain) project PID2022-138263NB-I0 (AEI/FEDER, UE). AGV is funded by “la Caixa” Foundation (ID 100010434) and the European Union's Horizon 2020 research and innovation programme under the Marie Sklodowska-Curie grant agreement No 847648, with fellowship code LCF/BQ/PI23/11970027. AGV and JSP are supported by projects  PID2022-136224NB-C21 (MICIU), 2021-SGR-00249 (Generalitat de Catalunya) and CEX2024-001451-M (ICCUB). JSP acknowledges participation in the COST Action CA23130 ``Bridging high and low
energies in search of QG'' (BridgeQG). The authors also acknowledge the participation in the COST Action CA21136 ``Addressing observational tensions in cosmology with systematics and fundamental physics'' (CosmoVerse).


\bibliographystyle{apsrev4-1}
\bibliography{references}

\end{document}